\documentclass[sigconf]{acmart}

\newcommand{\UMBC}{UMBC} 
\newcommand{\coursename}{HCC 629}
\newcommand {\change}[1]{{\textcolor{black}{#1}}}
 
\usepackage[normalem]{ulem}

\AtBeginDocument{%
  }

\copyrightyear{2026}
\acmYear{2026}
\setcopyright{cc}
\setcctype{by}
\acmConference[CHI '26]{Proceedings of the 2026 CHI Conference on Human Factors in Computing Systems}{April 13--17, 2026}{Barcelona, Spain}
\acmBooktitle{Proceedings of the 2026 CHI Conference on Human Factors in Computing Systems (CHI '26), April 13--17, 2026, Barcelona, Spain}
\acmDOI{10.1145/3772318.3790691}
\acmISBN{979-8-4007-2278-3/2026/04}

\begin{document}


\author{Kaoru Seki}
\authornote{Both authors contributed equally to this research.}
\email{kaoru2020@umbc.edu}
\orcid{0009-0004-2599-2787}
\affiliation{%
  \institution{University of Maryland, Baltimore County}
  \city{Baltimore}
  \state{MD}
  \country{USA}
}

\author{Manisha Vijay}
\authornotemark[1]
\email{mvijay1@umbc.edu}
\affiliation{%
  \institution{University of Maryland, Baltimore County}
  \city{Baltimore}
  \state{MD}
  \country{USA}
}

\author{Yasmine Kotturi}
\email{kotturi@umbc.edu}
\affiliation{%
  \institution{University of Maryland, Baltimore County}
  \city{Baltimore}
  \state{MD}
  \country{USA}
}

\title{Participatory, not Punitive: Student-Driven AI Policy Recommendations in a Design Classroom}

\begin{abstract}
Generative AI is reshaping education, yet most university AI policies are written without students and focus on penalizing misuse. 
This top-down approach sidelines those most affected from decisions that shape their everyday learning, resulting in confusion and fear about acceptable use. 
We examine how participatory, student-driven AI policy design can address this disconnect. 
We report on a three-part workshop series in a graduate design course at a minority-serving university in the U.S., where two student leaders facilitated discussions without faculty present. 
Eight participants shared candid accounts of their AI use, co-authored ten policy recommendations, and visualized them in a zine that circulated across campus. 
The resulting policies surfaced concerns absent from top-down governance, such as the double standard of requiring students to disclose or abstain from AI use while faculty face no such expectations. 
We argue that engaging students in AI governance carries value beyond the resulting policies, and offer transferable strategies for fostering participation across disciplines---a model for calling students \emph{in} rather than calling students \emph{out}.
\end{abstract}

\keywords{generative AI, participatory governance, higher education, student perspectives, design pedagogy}
\begin{teaserfigure}
  \includegraphics[width=\textwidth]{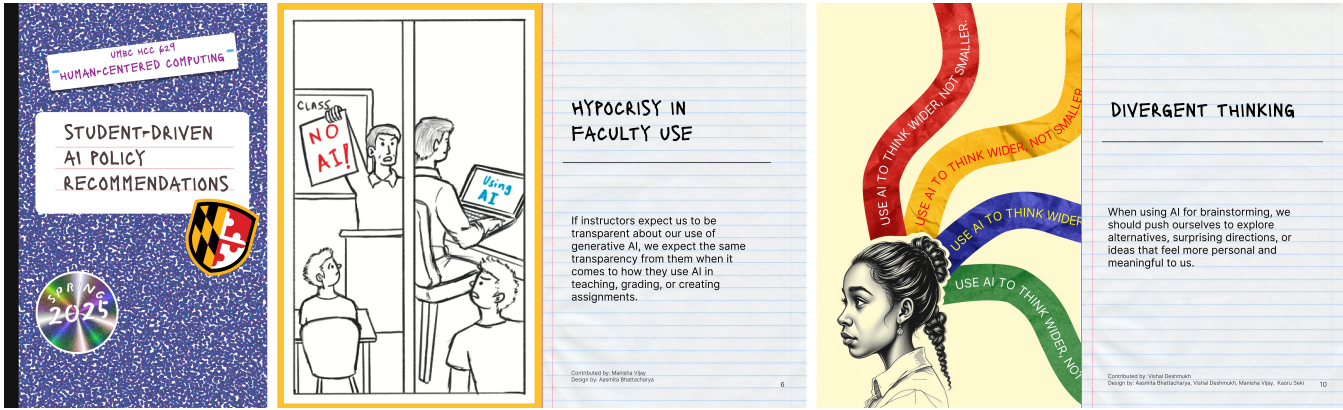}
  \caption{Ten graduate design students at a minority-serving university co-authored 10 policy recommendations through student-led workshops and follow-up interviews, then visualized them in a zine—a shareable, DIY booklet. Student-led discussions revealed fresh perspectives on existing policies and surfaced considerations often missed in top-down AI mandates. The initial zine pages shown above, and the full zine is viewable at \url{https://ykotturi.github.io/zines/}}
  \label{fig:teaser}
\end{teaserfigure}

\maketitle
\section{Introduction}
\label{sec:intro}
Generative artificial intelligence (AI) is rapidly reshaping teaching and learning practices across universities.
Although not the first technology to disrupt higher education~\cite{marshall2024reshaping}, generative AI technologies capable of producing text, images, and other media~\cite{sengarGenerativeArtificialIntelligence2025} have gripped administrators, faculty, and students alike, all racing to manage its challenges and opportunities~\cite{smolansky2023educator}.
Challenges are far-reaching: from questions of students' academic integrity~\cite{birks2023linking} and over-reliance \cite{zheng2025students}, to reductions in independent learning and weakened skill transfer \cite{lee2025impact}.
Overwhelmingly, responses to these challenges have led to university policies that emphasize the prevention of misconduct and plagiarism~\cite{alsharefeen2025examining}, which are often framed in punitive terms rather than pedagogical support~\cite{francis2025generative, luo2024critical}.

Across universities nationwide, one pattern holds in AI policy discourse: students' perspectives are largely absent~\cite[p.134]{bowenTeachingAI2024}.
And yet, students are the primary group affected by universities' AI policies~\cite{zhengChartingFutureAI2024, puHowCanWe2025}.
As noted in recent calls for a ``participatory turn in AI''~\cite{delgadoParticipatoryTurnAI2023}, omitting key stakeholders' perspectives and experiences from governing decisions lessens the efficacy of these decisions, and may even result in harmful outcomes due to exclusionary thinking ~\cite{gautam2024reconfiguring}.
Students bring nuanced perspectives into how AI is \textit{actually} used in everyday coursework---insights often invisible in top-down policy discourse and enforcement ~\cite{kuoPolicyCraftSupportingCollaborative2025}.
For instance, students are enthusiastic about using AI to improve personalized learning opportunities, self-reflection, and overall efficiency~\cite{kizilcecPerceivedImpactGenerative2024}.
In addition, students are concerned with AI extending instructors' surveillance capabilities~\cite{gorichanaz2023accused}, fueling incorrect accusations of AI plagiarism~\cite{giray2024problem}, and eroding trust in student-teacher relationships~\cite{hanStudentsPerceptionsExploring2025}.
Despite the richness and importance of these perspectives, there is little evidence that they are taken into account when shaping generative AI policy~\cite{weichertAssessingComputerScience2025}.

Therefore, in this paper, we take up calls to incorporate student perspectives in AI policies ~\cite{bowenTeachingAI2024,kizilcecPerceivedImpactGenerative2024, chanComprehensiveAIPolicy2023, puHowCanWe2025, hanStudentsPerceptionsExploring2025}.
We frame students as ``lead users''~\cite{von2006democratizing}, or early adopters of AI technologies.
In other words, we position students as experts in their own lived experiences with generative AI---offering insights that faculty and administrators can ultimately learn from when determining how (and how not) to integrate these technologies into the classroom.
In doing so, we work towards student-driven policy recommendations, where all stakeholders’ perspectives, especially those most affected by the policies, should have the ability to mold such policies~\cite{kuoPolicyCraftSupportingCollaborative2025}. 

We conducted a three-part, participatory design workshop series with follow-up interviews with 10 students from a graduate-level design studio course at University of Maryland, Baltimore County (UMBC): a minority-serving public institution in the mid-Atlantic region of the United States. 
To address inherent power dynamics between students and faculty, we implemented specific procedural safeguards: workshop activities were student-led to foster open dialogue without faculty oversight, and the research team ensured that faculty had no access to raw transcripts.
By grounding workshops in a class that all participants had recently taken,  discussions and recommendations were based on recent experience~\cite{kuoPolicyCraftSupportingCollaborative2025}, making for higher quality recommendations as students do not typically have prior policy writing experience~\cite{matthews2018engaging}. 

\begin{figure*}[ht]
  \centering
  \includegraphics[width=\textwidth]{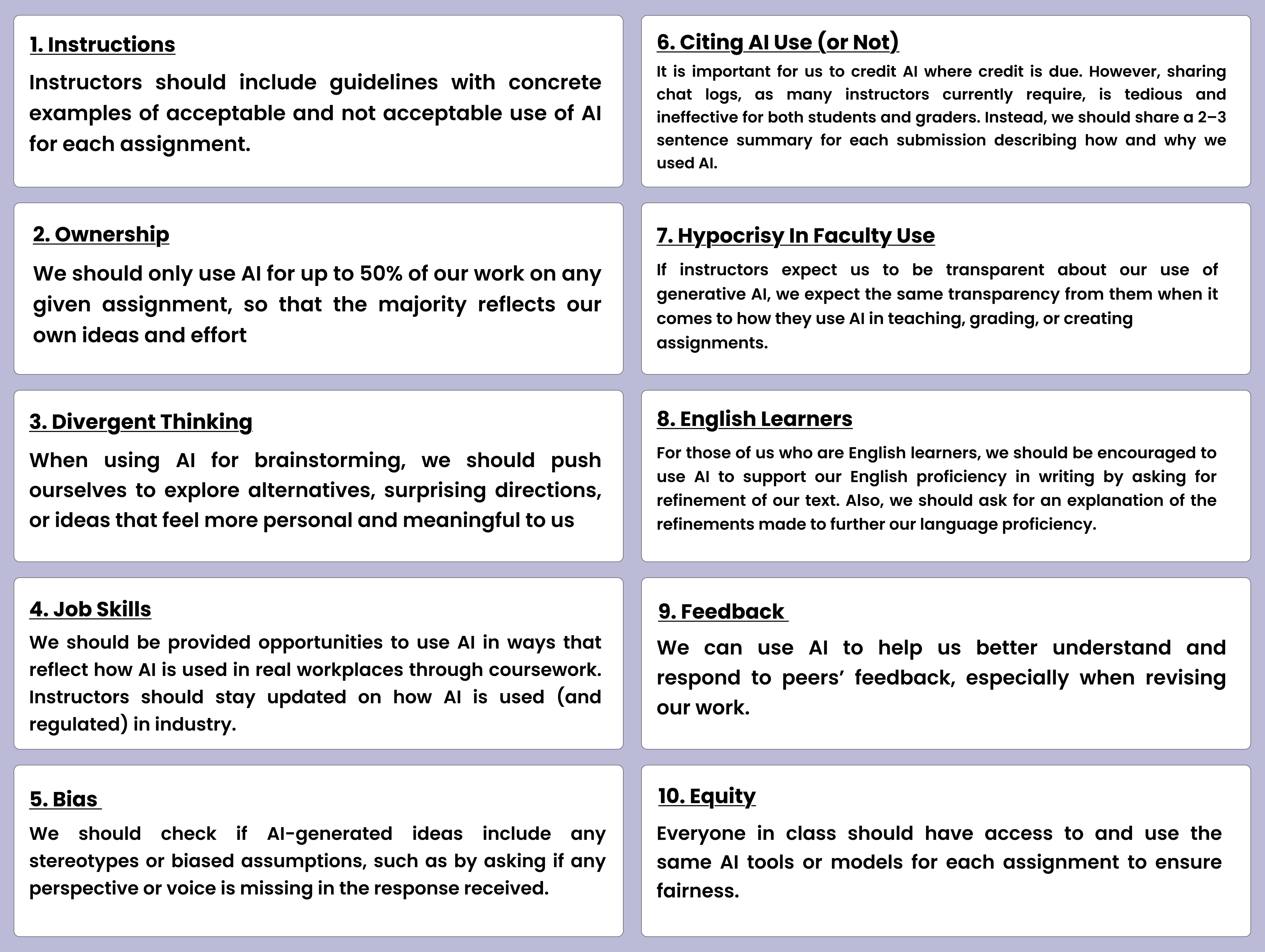}
  \caption{Ten student-drive AI policy recommendations derived from our three-part design workshop series}
  \Description{This table presents ten AI policy recommendations authored by students during a three-part workshop series. Each recommendation is phrased as a principle and paired with a brief explanation. The policies address: (1) Instructions: instructors giving clear assignment-specific AI guidelines, (2) Ownership: limiting AI contributions to no more than 50\% of a student’s work, (3) Divergent Thinking: encouraging divergent thinking when using AI, (4) Job Skills: integrating job-relevant AI skills into coursework, (5)Bias: checking for bias in AI outputs,  (6) Citing AI use (or Not): citing AI with short summaries instead of full chat logs, (7)Hypocracy in Faculty Use: faculty transparency in AI use (8) English Leaners: supporting English learners through AI-assisted refinement and explanation, (9) Feedback: using AI to support peer feedback, and (10) Equity: ensuring equity by providing access to the same AI tools for all students.}
  \label{fig:policy-table}
\end{figure*}

After an iterative design cycle (brainstorming, drafting, applying, and refining policies), we report on 10 resulting policies (See Figure~\ref{fig:policy-table}).
Policies were embedded in a student-authored zine---a DIY booklet~\cite{licona2012zines, duncombe1997notes} used as an artifact and method to promulgate information---which was circulated widely to support student-driven discourse across the campus (See Figure ~\ref{fig:teaser}), as well as archived in \UMBC's Library\footnote{A digital version of the zine archived by the UMBC Library can be found here: \newline 
\url{https://lib.guides.umbc.edu/c.php?g=1475372&p=11187429}}.
We view this distribution strategy as a form of participatory infrastructuring ~\cite{dantecInfrastructuringFormationPublics2013}:
by archiving the zine and circulating it through physical and digital channels, we moved beyond the workshop series as a singular event, instead creating lasting resources that allow student voices to persist in institutional memory and continue shaping policy long after the initial workshop activities concluded.

In addition, through thematic analysis and triangulation of workshop dialogue, interviews, and artifacts (e.g., zines, design activities), we find how participants, when provided the environment to speak candidly, gradually disclosed more layered accounts of AI use, such as when and why they used AI that contradicted existing course policies. 
While student-authored policy recommendations reflect students' leniency towards AI use writ large~\cite{10.1002/pra2.1277, sah2025generative}, they also capture the nuanced, lived perspectives of those most directly implicated in governance decisions.
Importantly, we observed how the process of authoring policies shifted participants' practice towards more intentional, reflective AI use, suggesting that the process of participation may be equally important to the outputs.

Together, this paper makes three contributions. 
First, we contribute an empirical account of students' candid AI practices in a graduate design studio—including misuse, policy gray areas, and uneven enforcement. 
Second, we contribute a transferable model for student-driven AI policy design that combines faculty-free workshops, student-led facilitation, and zine-based participatory infrastructuring to lower barriers to candid expression and support participatory governance in higher education. 
Third, we contribute ten actionable, design-oriented policy recommendations authored by students, situated within the broader landscape of institutional AI policies---highlighting where student perspectives align with, extend, or diverge from prevailing administrative approaches. 
We reflect on both the challenges and the benefits of student-driven policy recommendations, arguing that the participatory process itself---calling students in as co-authors of governance rather than subjects of regulation---carries value beyond the policies it produces.



\section{Related Work}
We draw on three bodies of scholarship to inform this work: how generative AI is (and is not~\cite{pang2025shaping}) reshaping the classroom in higher education, how diverse participation improves efficacy and trust in AI governance, and how participatory infrastructuring enacted through zine-making offers a framework to actualize participatory aims of universities' AI governance.

\subsection{Generative AI in the (Design) Classroom}
Generative AI has quickly become a contested presence in higher education, continuing a long lineage of technological intervention in the classroom---from calculators in the 1970s to MOOCs in the 2010s---that have disrupted teaching and learning. 
Unlike earlier technologies, however, the scale and scope of automation possible with generative AI uproots and nullifies many of the safeguards educators rely on to ensure adequate learning~\cite{luo2024critical}.
For instance, use of text-generation AI tools (such as ChatGPT or Claude), can reduce students' opportunities to experience ``desirable difficulties''--or the productive cognitive struggle that causes long-term retention and skill transfer~\cite{bjork2011making}.
As a result of this decreased cognitive engagement, students who use generative AI technologies in unstructured ways may not develop essential skills: critical thinking, problem-solving, and creativity~\cite{kosmyna2025your, gerlich2025ai}. 
Given these risks, universities have been compelled to regulate their students' use of AI technologies~\cite{jin2025generative, luo2024critical}, resulting in a cacophony of regulatory attempts~\cite{moorhouse2023generative}.

To understand this capricious policy environment, recent scholarship categorizes institutional responses along a spectrum from restrictive to experimental~\cite{sumilong2025instructional}. 
On the restrictive end, some universities have implemented a blanket ban on AI use, like Sciences Po in Paris, France, which prohibits generative AI use without explicit authorization under threat of expulsion~\cite{Hristova2025GenerativeAI}.
On the experimental end, initiatives like the University of Florida's ``AI Across the Curriculum'' ~\cite{Southworth2023AIAcross} and Ohio State's ``AI Fluency'' initiative~\cite{osu2025fluency}, claim that all of their graduates (not just their computing and engineering students) will be trained with ``essential AI skills'' such as using ``AI tools to accomplish specific goals in the field of study, and critically assess outputs for accuracy and relevance to the task''~\cite{Southworth2023AIAcross, osu2025fluency}. 
Arizona State University's recent partnership with OpenAI takes this experimentation a step further to launch ``innovation challenges'' inviting students to co-create use cases for what AI use may be beneficial~\cite{ASU_PersonalizedLearning_2025}. 
And of course, there are many institutions in between these two extremes, such as at University of Maryland, Baltimore County (UMBC)---the site of this study---where recent AI policy guidelines recommend that instructors update their syllabi to signal to students what level of AI use is permitted in class:  ``Green light'' (unrestricted), ``Yellow light'' (some restrictions), and ``Red light'' (prohibited). 
 
Across these myriad approaches to regulate student use of AI, from the lenient to punitive, key problems emerge.
First, there is often little explanation for the rationale behind the policies. Implementing AI policy without articulating the pedagogical reasoning, whether concerns around learning outcomes, academic integrity, and skill developments, may lead students to see it as arbitrary practices of authority rather than thoughtful pedagogical decisions~\cite{steingut2017effect}. 
Without meaningful rationale, students are less likely to internalize the policy's value, resulting in diminished engagement, autonomous motivation, and performance ~\cite{deci1994facilitating, aelterman2019correlates}.

Another key challenge with current approaches to AI governance in higher education is a lack of reciprocity: while students must disclose their usage and/or are prohibited from using AI, instructors are not held to the same standard, and can use AI for teaching or grading without similar disclosure rules~\cite{luo2025does}. 
This asymmetry reinforces power imbalances and reduces trust between students and faculty~\cite{luo2025does}. 
Finally, another challenge includes the enforcement of policies: making sure that students only use AI within the bounds of allowed use. 
The predominant approach to address this challenge comes laden with additional problems: AI detection tools, though intended to prevent misconduct, introduce serious risks of false accusations~\cite{giray2024problem, gorichanaz2023accused}. 
Unlike plagiarism systems such as TurnItIn.com that compare work against known sources~\cite{turnitin2025}, AI detectors infer authorship probabilistically, making accusations far more uncertain and contestable~\cite{erol2025can}. 
These issues suggest a fundamental limitation of top-down AI policy design that excludes students—the most affected stakeholders—from its implementation.

\subsubsection{Generative AI in the Design Studio: Opportunities and Risks}
While generative AI is reshaping higher education broadly, this paper focuses on its impact on design pedagogy. 
Design studios are a salient site for study because they rely on open-ended inquiry, deliberate practice, and iterative processes~\cite{li2024user}---all practices permeable to AI's multimodal capabilities. 
Empirical studies paint a complex picture of AI in design which includes both opportunities and risks
towards opportunity: generative AI can support early phases of conceptual design—problem framing and ideation~\cite{chenHowGenerativeAI2025}. 
Design students describe large language models (LLMs) as a ``second mind'' for externalizing and iterating ideas~\cite{wanItFeltHaving2024}, aligning with views of AI as collaborator rather than tool~\cite{sarkar2023enough}. 
Additionally, hybrid systems merging physical prototyping with AI open new avenues for low-fidelity exploration~\cite{zhangProtoDreamerMixedprototypeTool2024}, and antagonistic AI can provoke divergence rather than premature convergence during ideation~\cite{liuSmartErrorExploring2024, cai2024antagonistic}.

On the other hand, risks persist: AI-assisted ideation can reduce variation of ideas~\cite{wadinambiarachchiEffectsGenerativeAI2024}, as prompting practices can homogenize outputs when users converge on popular modifiers~\cite{oppenlaender2025prompting}.
This homogeneity is then exacerbated by generative models' statistical regularities in training data, which steer creators toward shared stylistic outcomes~\cite{oppenlaender2025prompting, wadinambiarachchiEffectsGenerativeAI2024}.
In other words, even with careful prompting techniques that anticipate AI's homogeneous tendencies, the underlying technology that powers generative AI technologies can undermine these end-user attempts to course correct. 
In addition, issues of consent, credit, and compensation remain unresolved: AI models are trained on creatives' works without consent~\cite{kyi2025governance}, spurring calls for alternative ownership paradigms~\cite{polimetla2025paradigm}.
Our study contributes by examining how design students themselves articulate these tensions and propose ways to mitigate risks and pursue opportunities. 
Given that students are rarely asked to be involved in such governance discussions, such an investigation requires a participatory approach, as discussed in the next section.

\subsection{The Participatory Turn in AI}
Recent HCI scholarship has focused on the ``participatory turn'' in AI, where decision-making for AI systems' design and governance should not be limited to the creators, but also includes those who are impacted by such technologies~\cite{delgadoParticipatoryTurnAI2023}. 
This repositioning emphasizes the importance of acknowledging stakeholders' lived experience and involving them not only in giving feedback on systems but also in co-setting agendas and informing what is built in the first place~\cite{gautam2024reconfiguring, birhane2022power}.
In addition to improved efficacy, participatory approaches to AI governance may also serve a trust-repair function.
For instance, in higher education where AI infiltration has damaged student-faculty relationships~\cite{gorichanaz2023accused, Petricini2025}, participatory approaches to AI governance can help to repair breakdowns by calling students \emph{in} rather than calling students \emph{out}.
Students-as-partners scholarship shows that co-creation shifts classroom dynamics from hierarchical to collaborative, with benefits for both student engagement and learning outcomes~\cite{bovill2011cocreators, mercermapstone2017systematic}.

Importantly, participatory AI can slide into tokenism without intentional steps to anticipate and course correct~\cite{birhane2022power}. 
Delgado et al. highlight how many participatory AI efforts remain consultative rather than empowering~\cite{delgadoParticipatoryTurnAI2023}.
Pushing past tokenism requires moving participation upstream, sharing decision authority, and creating safe conditions for dissent. 
To do this successfully requires a clear-eyed understanding of the barriers to participatory design approaches: it is time and resource-intensive, may not scale easily across institutional contexts, and risks placing burdens on participants asked to perform tasks without formal training~\cite{vines2013configuring, bossenScalingPD2022}. 
In the context of participatory governance in the classroom, tools like PolicyCraft~\cite{kuoPolicyCraftSupportingCollaborative2025} mitigate these challenges with ``case-grounded deliberation,'' structuring policy discussion around concrete scenarios, seeded by facilitators, to make governance decisions easier for non-experts, such as students. 

In this paper, we consider another approach to fostering student participation in AI governance, while acknowledging the challenges that come alongside fostering participation.
We consider power asymmetries between students and faculty, even in ostensibly participatory spaces, requiring deliberate structural safeguards~\cite{bovill2020cocreation}.
We frame students as ``lead users''---or early-adopters~\cite{von2006democratizing}---who have particular expertise from lived experience that instructors may lack, and design power-aware, student-led workshop activities in order to lower barriers to candid expression and resist tokenistic consultation.
Finally, we center student discourse around a zine-making activity in order to emphasize not only the resulting artifact, but also the importance of deliberation among students along the way (as described in the next section).

\subsection{Participatory Infrastructuring through Zine-Making}
In this work, we frame zines---small, self-published booklets~\cite{atton1999reassessment, duncombe1997notes}---as infrastructure: material media that circulate student perspectives across classrooms and institutions while catalyzing social formations and shared knowledge.  
Their low-cost, DIY format positions them as alternative media for grassroots circulation~\cite{mcnutt2021potential}, amenable to candid authorship and resistant to polish and performativity.
Zine-making has played a critical role in feminist, punk, and activist communities to amplify marginalized voices and foster counterpublics~\cite{licona2012zines, duncombe1997notes}. 
More recently, zines have been used as a participatory design method in HCI to surface marginalized perspectives~\cite{hayZineographyCommunityBasedResearchthroughDesign2024, harrington2019deconstructing}.
In contrast to top-down policy memos that reduce student agency to checkbox compliance~\cite{ghimire2024guidelinesgovernancestudyai}, zines may allow students to narrate their own experiences, reflect on trade-offs, and suggest alternatives. 
It is important to note that zine-making can be challenging for those unfamiliar with their format, but strategies such as detailed explanations, examples, and templates can ameliorate this~\cite{hayZineographyCommunityBasedResearchthroughDesign2024}.

We view zine making through the lens of participatory infrastructuring, which emphasizes creating sociotechnical conditions that allow publics to form---not pre-defined stakeholder groups, but emergent collectives constituted through shared concern and active engagement~\cite{dantecInfrastructuringFormationPublics2013}. 
These publics participate not merely as respondents, but as co-constructors of the problem space itself~\cite{dantecInfrastructuringFormationPublics2013, asadTapMakeThis2017}.
Through a participatory infrastructuring lens, we frame student-led zine-making as collective world-building: students shifting from a fixed stakeholder group to a reflexive public capable of intervening in institutional discourse.
Student-centered approaches to policy articulation are especially important in higher education, where classroom authority is traditionally asymmetrical. 
Scholars of critical pedagogy argue that shifting power toward students can improve learning outcomes and foster civic capacities~\cite{hess2007collaborative, moreno2005sharing, shor1996students}.
Student-driven policies may be more likely to be respected and internalized when they reflect lived realities~\cite{bowenTeachingAI2024}; recent work by Pu et al. demonstrates the value of engaging students in co-designing AI guidelines grounded in their concerns~\cite{puHowCanWe2025}. 

\begin{figure*}[ht]
  \centering
  \includegraphics[width=\textwidth]{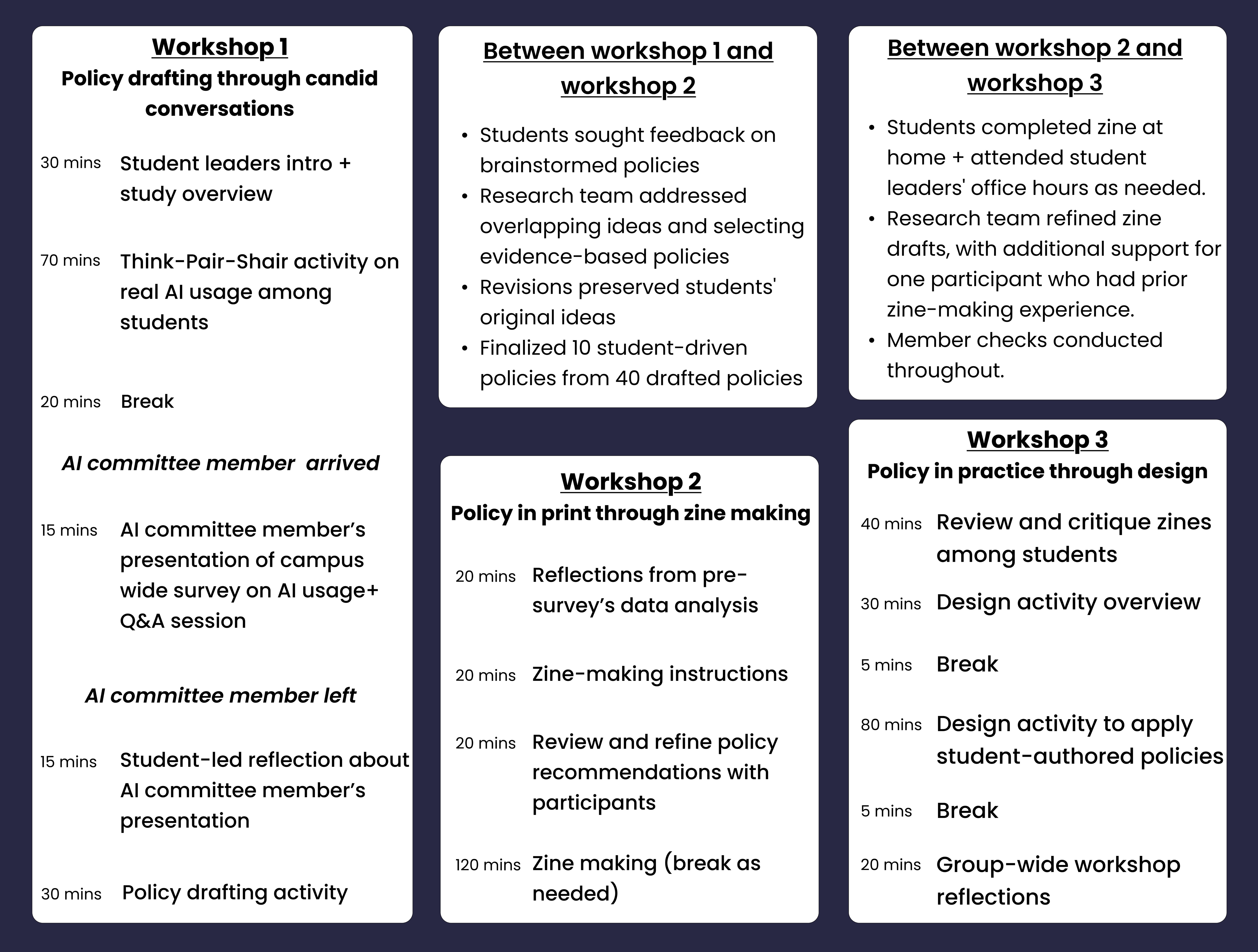}
  \caption{Overview of the three-part workshop series.}
  \Description{This figure presents a three-workshop participatory design process for student-driven AI policy creation, shown as six labeled panels on a purple background. Workshop 1, titled “Policy drafting through candid conversations,” outlines a 4-hour session with timed activities: student leader introduction and study overview (40 minutes), a Think-Pair-Share on real AI usage among students (70 minutes), a break (10 minutes), a campus AI committee member’s presentation of survey results with Q and A (30 minutes), student-led reflection on the presentation (30 minutes), and a policy drafting activity (60 minutes).A middle panel describes activities between Workshop 1 and Workshop 2, including student feedback seeking, research team synthesis of overlapping ideas, preservation of students’ original ideas, and finalization of 10 student-driven policies from 40 drafts. Workshop 2, titled “Policy in print through zine making,” includes reflections on pre-survey data (20 minutes), zine-making instructions (20 minutes), review and refinement of policy recommendations (20 minutes), and zine creation (120 minutes, with breaks as needed). Another panel describes activities between Workshop 2 and Workshop 3, including students completing zines at home, attending office hours as needed, research team refinement of zines with additional support for one participant, and ongoing member checks. Workshop 3, titled “Policy in practice through design,” includes reviewing and critiquing zines (40 minutes), a design activity overview (30 minutes), a short break (5 minutes), an extended design activity applying student-authored policies (80 minutes), another short break (5 minutes), and a final group-wide reflection (20 minutes).}
  \label{fig:workshop-series}
\end{figure*}

\section{Methods}
\label{sec:methods}
We draw on community-based participatory design~\cite{harrington2019deconstructing} and conducted a three-part, in-person design workshop series to facilitate policy articulation, application, and iteration.
This workshop series was led by two student research assistants, who are joint first authors of this paper.

\subsection{University setting}
This study was conducted at University of Maryland, Baltimore County (UMBC): a minority-serving, public university in the U.S. mid-Atlantic. 
To contextualize AI use among UMBC's student body relative to other universities, a campus survey indicated slightly slower AI uptake relative to national reports: 27\% reported using AI in coursework (vs. 42\% in a national study~\cite{TytonPartners_TimeForClass2025}); 73\% had not taken a formal AI course and AI use skewed heavily towards STEM majors~\cite{mcdonaldGenerativeArtificialIntelligence2025, ft2024}. 
Therefore, UMBC represents a context where formal AI policies and institutional support structures are still early-stage.

\subsection{Participants and recruitment}
Participants were recruited from the last author's graduate design studio course, \coursename: Fundamentals of Human-Centered Computing, taught the preceding semester before the study (Fall 2024). 
This is a required course for the master's and Ph.D. students in the department's HCI specialization, and focuses on fundamental design principles and human factor concepts to guide effective user interface design.
Of 21 enrolled students, 10 completed a screener with questions such as ``Describe how you used genAI in \coursename?'' and ``At \UMBC, how do existing AI policies support/hinder your learning?'' Eight students were selected to vary course performance~\cite{johnstonStudentPerspectivesUse2024}, AI use, schedules, and degrees (e.g., HCI, information science, software engineering).
Informed consent from all participants was acquired, and all ethical procedures adhered to the standards of the university institutional review board, as per the approved IRB protocol by which this research was conducted.
Workshop participants were compensated \$20/hour (11 hours; \$220 total for each workshop participant for the three-part workshop participation).
Following the workshop series, we conducted five follow-up interviews: three with workshop participants and two with students who declined workshop participation due to dissenting views on AI or scheduling conflicts. 
Workshop participants who completed follow-up interviews were compensated an additional \$20; interview-only participants were compensated \$20 for one hour of their time (described in detail in Section~\ref{sec:interviews}).
In total, 10 students contributed: 8 workshop participants and 2 interview-only participants (degrees: 7 HCI, 2 Information Science, 1 Software Engineering; 9 master's, 1 Ph.D.).
To protect identities given public zine authorship, we report only these high-level demographics.

\subsection{Three-part workshop series}
A total of three workshops were held on Fridays: March 28, April 11, and April 25, 2025, from 12:00--3:00pm; lunch was provided.
See Figure~\ref{fig:workshop-series} for an overview of workshop structure and duration.
Each workshop was held in a design studio classroom with modular tables, chairs, and whiteboards to support collaborative activities.
Prior to the workshop series, participants completed a pre-workshop survey to capture their motivations to participate, and initial opinions, skills, and uses of generative AI in the graduate design studio course (and beyond).
Students' top two motivations to participate included learning how to use AI more ethically, and having their perspectives taken into consideration in AI policy creation. 

\begin{figure}[ht]
  \centering
  \includegraphics[width=0.45\textwidth]{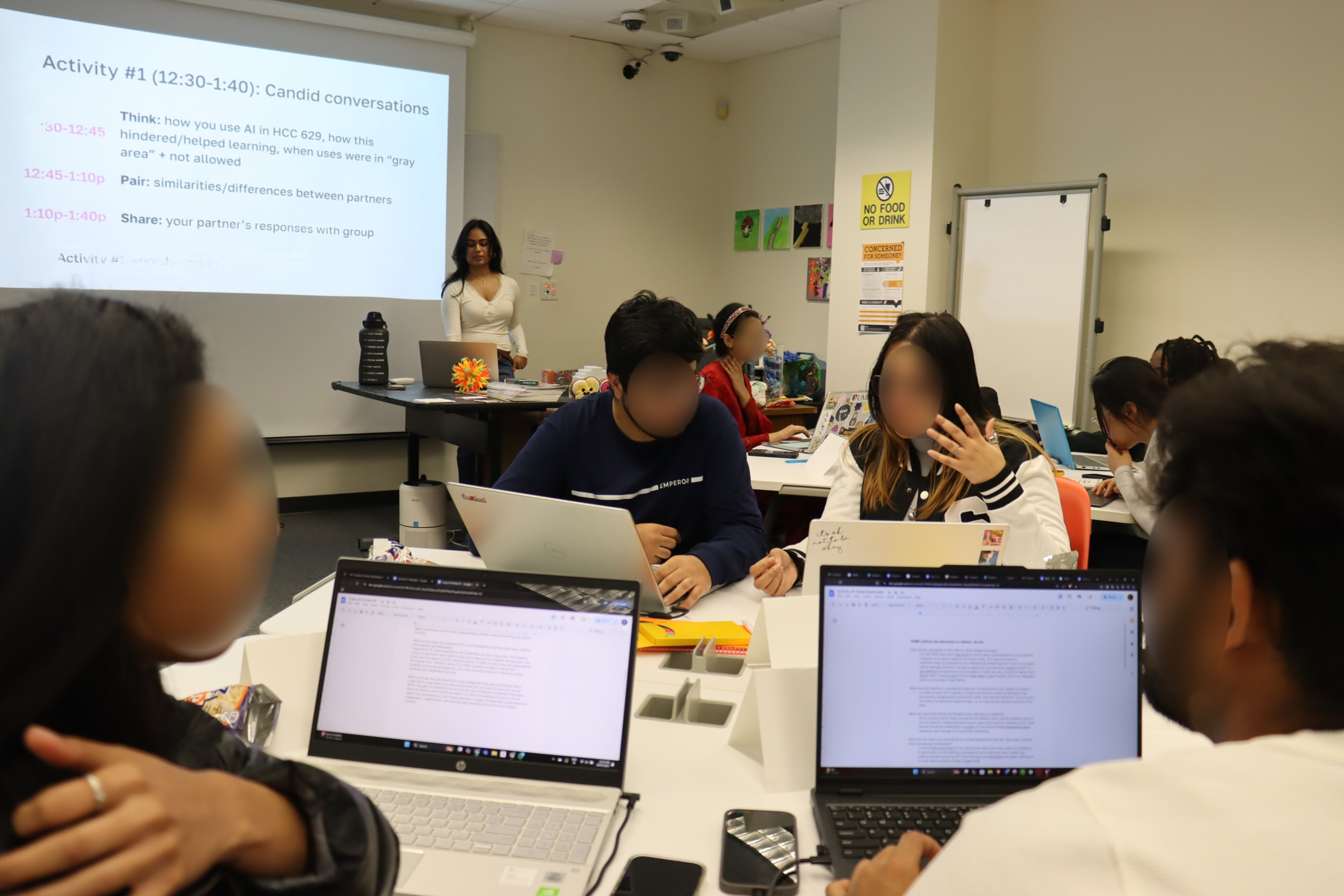}
  \caption{Workshop 1 focused on policy drafting through candid conversations of AI use. Students exchanged reflections on their use of AI tools in a design course the preceding semester.}
  \Description{Photograph of Workshop 1. Students are seated in pairs and small groups, leaning toward one another as they talk and exchange perspectives. The activity shown is the 'share' phase of a think–pair–share exercise, where participants are articulating their reflections on classroom AI usage.}
  \label{fig:workshop}
\end{figure}

\subsubsection{Workshop 1: Policy Drafting through Candid Conversations}
\label{sec:workshop1}
Workshop 1 focused on supporting candid discussions among students as a launching point to brainstorm policy recommendations.
Given the sometimes punitive environment surrounding generative AI use, we took steps to create an environment where students felt comfortable disclosing use, voicing concerns, and sharing guidance with each other. 
First, to ensure participants had a safe space to converse frankly and honestly, these conversations were driven by the two student leaders sans faculty. 
These two student leaders were graduate peers of the participants rather than instructors or teaching assistants; they held no formal evaluative role (e.g., grading, advising, or assessment responsibilities) in relation to the students who took part in the workshops or follow-up interviews.
Second, the faculty PI did not have access to the recordings nor the raw transcript, and could only view a de-identified version of the transcripts (this was explained repeatedly to participants throughout the workshop).
Finally, student leaders reminded participants they have essential lived experience with AI tools, and that the study aimed to learn from them (See Supplemental Materials for workshop slides and all other workshop materials).

\textbf{Think-Pair-Share to Ground Discussions of AI Use.} During the first workshop, participants engaged in a ``Think-Pair-Share''~\cite{lyman1981responsive} where they reflected on their use of AI (e.g., how their use fell into a ``gray area'' of acceptability or how they used AI in ways that violated their courses' AI policy), compared these experiences with a peer, and then partook in a group-wide discussion that elevated key themes.
Students were encouraged to ground their reflections in recent experience.
To help students situate their experiences within a broader campus context, a guest speaker from the university's Ad Hoc Committee for AI in the Classroom shared findings from a university-wide survey of 200 students, including data on faculty adoption and dissent. 
The student leaders then facilitated a discussion among workshop participants (See Figure~\ref{fig:workshop-series}).

\textbf{Policy Drafting.} Student leaders then introduced participants to the policy drafting activity. 
Following Sanders and Stappers' recommendations for researcher-as-facilitator in participatory design~\cite{sanders2008co}, we treated AI policy design as a domain in which students needed structured entry points rather than a blank slate (based on participants' responses from the pre-workshop survey). 
As a starting point, student leaders presented 24 potential policy topics, drawn from \coursename's learning objectives (divergent thinking, feedback), related AI education literature (e.g., agency and equity~\cite{kotturiDeconstructingVeneerSimplicity2024}, critical thinking~\cite{Satyanarayan2024Intelligence}, ownership, group work, and social support~\cite{bowenTeachingAI2024}), ongoing AI policy conversations at \UMBC~(e.g., AI literacy, grading and rubrics, academic integrity), and the student leaders' own experiences (e.g., perceived \change{double standards} in faculty use, English learner considerations).
Participants were invited to add additional topics (See Supplemental Materials, Workshop 1 Slides, Slide \#34), and one participant did (i.e., Policy \#1 ``Instructions'').
This approach allowed us to seed the workshop with topics that were both pedagogically grounded and locally salient, giving novice policy writers clear, student-relevant entry points while treating the list as a sensitizing scaffold rather than an exhaustive or prescriptive agenda.

Each participant then selected five topics to draft policy recommendations for.
Participants were provided with structured prompts on an activity worksheet that guided both brainstorming and policy formulation.
Each policy topic---such as ``Divergent Thinking'' or ``Hypocrisy in Faculty Use''---was presented with a guiding question (e.g., ``How can generative AI encourage you to explore diverse ideas and inspire creative approaches to problem-solving?'').
Prompting questions helped translate abstract concepts into student-facing questions in everyday language that they could more easily understand how to map their experiences and make policy topics connect to their lived experience.
For each theme, worksheets led students through a process to identify a core challenge or tension, describe a relevant learning outcome or classroom scenario, and finally draft a concise policy statement using conditional phrasing and action verbs. 
This policy drafting activity resulted in 40 initial recommendations (8 participants, 5 prompts each).

\textbf{Between Workshop 1 and Workshop 2.} The research team convened three times to refine the 40 initial policy recommendations. 
While the core sentiment of the student recommendations was preserved, the research team rephrased statements to consolidate vague rhetoric, such as ``do not use AI to complete your work.''
Widely repeated suggestions, such as many policies focused on academic integrity, were consolidated (See Table~\ref{fig:policy-table}, Policy \#6 ``Citing AI Use (or Not)'').
The research team contacted participants with clarifications to preserve student intent.
The authors reviewed the initial recommendations with related literature to ensure policies were backed by evidence, as well as revising for clarity and novelty.
For instance, a participant recommended that ``students should only use genAI at later stages of the design process to ensure ideas are their own.''
However, generative AI shows potential for supporting early ideation~\cite{hsiaoStudyApplicationGenerative2024a}. 
This synthesis process ultimately resulted in a curated set of 10 student-driven policy recommendations (See Table~\ref{fig:policy-table} for this list of final policy recommendations).

\begin{figure}[ht]
  \centering
  \includegraphics[width=0.45\textwidth]{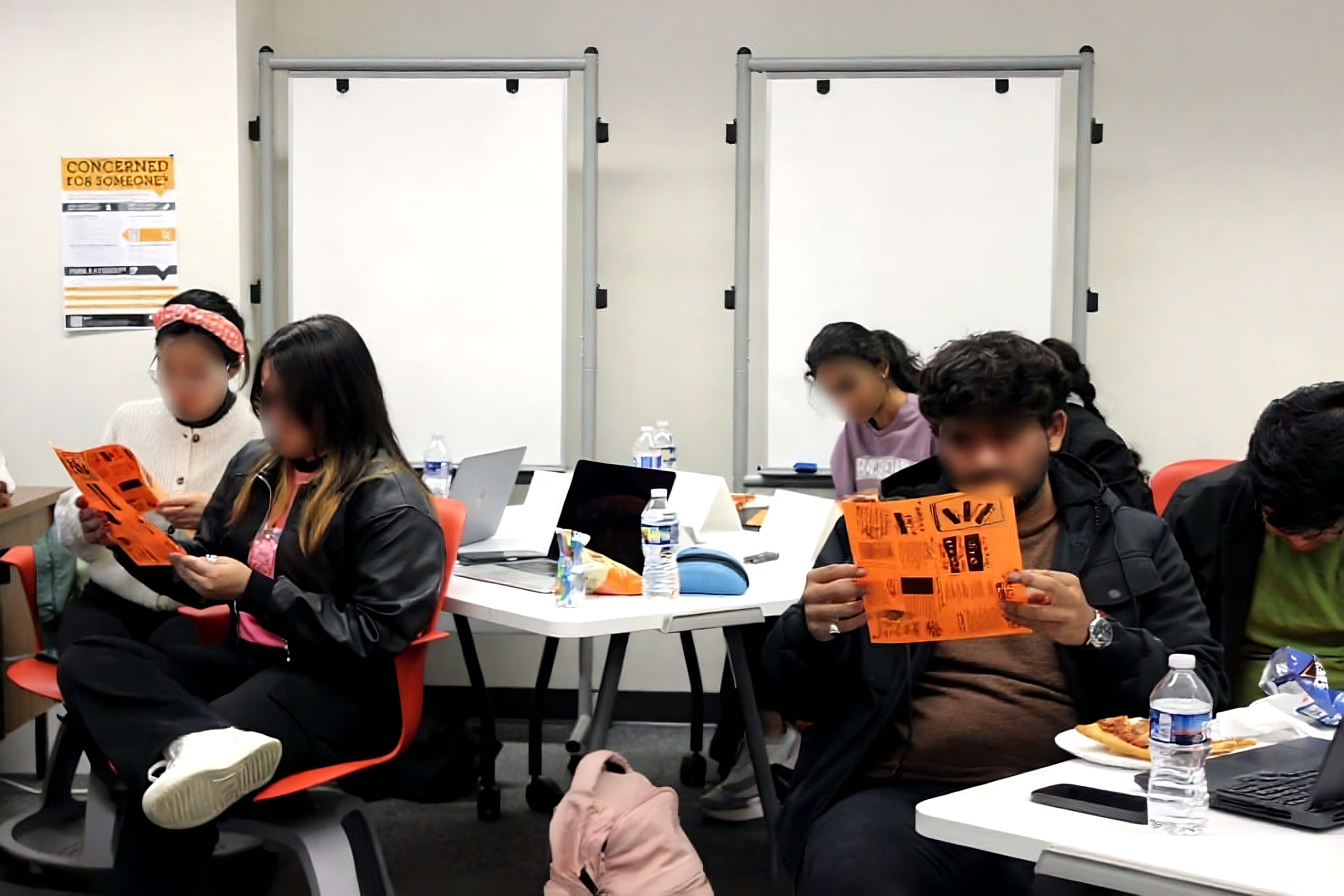}
  \caption{Workshop 2 focused on zine-making. Students flip through several exemplar zines while learning about the historical context of zine-making practices.}
  \Description{The photo shows a classroom workshop during the zine-making activity. Several students are seated around tables, each holding and reading bright orange handouts provided by facilitators as supplemental material. Laptops, snacks, and water bottles are on the tables. This moment depicts the instruction phase, when participants reviewed facilitator-provided materials before creating their own zines on AI policy.}
  \label{fig:workshop2}
\end{figure}

\subsubsection{Workshop 2: Policy in Print through Zine Making}
\label{sec:workshop2}
After the research team and participants went through the consolidated and refined policy recommendations, participants started the zine-making process by choosing two of the policies to visualize.

\textbf{Introduction to Zine Making.} Given that most participants had never made a zine before, the research team took steps to onboard students to the zine-making process~\cite{hayZineographyCommunityBasedResearchthroughDesign2024}. 
Student leaders welcomed a visual arts professor at the university who leverages \textit{zines as research}~\cite{hay2022zineography} through their extensive experience of zine making (See Workshop 2 Slides 10--27 in Supplementary Materials).
The professor reviewed zine history~\cite{Pink2016Tracing, QueerMusicHeritage_LisaBen_ViceVersa}, then facilitated discussion on the relevance of zine making in this project: ``Why not just write a typical policy document outlining findings?''; ``Why take the time to create a visual representation of a policy?''; ``What legacy of zine making is relevant, given the student-driven nature of this workshop series?'' 
Students were challenged to think about what it means to present policies in the form of a zine, rather than typical policy formatting. 
The research team accommodated multiple approaches to creating zine pages based on preferences and strengths, to further accommodate participants' familiarity with zine making. 
Using provided templates, each participant created one digital page in Figma~\cite{Figma} and one physical page on paper.
Students were provided materials such as markers, tape, scrapbook and magazine cut-outs, scissors, and glue sticks. 
Digital tools such as an ASCII art generator~\cite{ASCIIArt}, the Iconify Figma plugin ~\cite{Iconify}, and generative AI tools such as DALL-E~\cite{Dalle3} were also provided.

\textbf{Iterations and Refinements of Initial Zines.}
While participants had two hours during Workshop 2 to begin zine making, most finished their pages the subsequent week. 
Participants were allowed to borrow any physical materials provided during the workshop to complete their analog pages. 
To support participants, student leads held office hours to provide feedback on creative choices and clarify instructions from Workshop 2. This in-between counseling occurred alongside refinement of the draft zines. 
The research team met five times to review drafts (See Supplemental Materials for the first draft of zines). 
Member checks were also conducted to garner students' feedback before sending to print~\cite[Ch.4]{charmaz2006constructing}.

\subsubsection{Workshop 3: Policy in Practice through a Design Activity}
\label{sec:workshop3}
In the final workshop, students applied their policies to complete a design activity that resembled their assignment structure from the design studio course.
Participants were asked to apply their policy recommendations to redesign generative AI interfaces to better support the design studio course's learning objectives.
Students selected from six generative AI tools spanning image/UI-based (V0~\cite{V0app}, UX Pilot~\cite{UXPilot}, UIzard~\cite{Uizard}) and text-based (ChatGPT~\cite{ChatGPT}, Claude~\cite{Claude}, Grammarly~\cite{Grammarly}) categories, ensuring coverage across all tools.
Participants completed their redesign on Figma~\cite{Figma}, with as many screens as they desired, aiming for higher fidelity redesigns. 

\textbf{Design activity to apply policies.} 
To support their design activity, participants choose from a list of \coursename's objectives to optimize their design for as well as a conceptual metaphor to provide additional structure: AI as a co-agent~\cite{Satyanarayan2024Intelligence}, AI as a sensor not a solution~\cite{Satyanarayan2024Intelligence}, AI as Feedback Generator, Personal Tutor, Learner, and Team Coach~\cite{Mollick2023AI} (See Supplemental Materials for full list of conceptual metaphors). 
Participants then redesigned the interface of [GenAI tool] by using the conceptual model of [Conceptual Model] to strengthen [\coursename's learning objective].
To support this reflection on policies in practice, participants completed a reflection worksheet, which included opportunities to revise policies.

\textbf{Expert evaluation of interfaces.}
The participants' redesigned interfaces (included in Supplemental Materials) were then assessed by five senior human-computer interaction faculty at the university, all of whom had taught the design studio course previously and were familiar with the course's learning objectives.
Interfaces were evaluated on a Likert scale of 1--3 (1-needs improvement, 2-adequate, 3-strong) with four criteria: learning objective, conceptual model, usability, and overall design quality. 
Interfaces were ranked by average score; top interfaces may be built in future offerings of \coursename.

\textbf{Publishing the zine.} The refining stages to produce a camera-ready zine---included in the Appendix (See Appendix~\ref{app:appendix})---extended well beyond the third workshop, until July 2025, when the zines were printed at a student-run printing service on the university's campus.
All students involved were invited to celebrate the completion of the zine at a ``zine party'' in October 2025, where they received copies and discussed the project and ricocheting campus conversations over a provided lunch.
The last author implemented these student-driven policies in the proceeding offering of the same course, providing an opportunity to evaluate downstream effects in a live classroom. 

\textbf{Open-sourcing workshop materials.} Workshop materials are open-sourced so that other faculty and students can run these activities in different departments (across \UMBC but also in other institutions)---see Supplemental Materials.

\subsection{Post-Workshop Interviews}
\label{sec:interviews}
After the workshop series, the authors conducted three follow-up interviews with workshop participants to understand how their thinking on the topics discussed had developed since the workshop took place. 
We also conducted two additional interviews with students who were enrolled in the design studio course but opted not to participate in the workshop series due to dissenting views on AI in order to investigate non-adopters' perspectives on generative AI~\cite{zhou2025attention}.
We followed a semi-structured interview protocol, with differing questions for workshop participants and non-workshop participants. 
Workshop participants were asked questions such as: 
\emph{How has your use of AI in coursework stayed the same or changed since participating in this workshop series?}, and
\emph{Have you had any conversations about the workshops, zine, policies, etc. with your peers or instructors?}.
Those who did not participate in the workshops were asked questions such as: \emph{What were your concerns with participation in the workshop?} and
How can we ensure students with dissenting views of AI are heard at the policy level?
Interviews were conducted in person or online via WebEx for one hour and interview participants were compensated \$20 (on top of the \$220 compensation for those who also participated in workshops). 
We acquired informed consent from interview participants, and all data collected from the interviews were securely stored following the university's institutional review board's protocol for protecting research data.


\subsection{Data Analysis}
Audio recordings from workshops and interviews were transcribed with Otter.AI~\cite{Otter.ai} and manually checked for errors. The joint first authors corrected transcripts and de-identified all materials before sharing them with the faculty PI.

We conducted thematic analysis~\cite{braun2006using, braun2022thematic}, where all three authors participated in analysis.
All authors collaboratively coded the first workshop transcript and created a living codebook.
Then, the two joint first authors led the coding of all remaining transcripts (both interviews and workshops), and the faculty PI joined for codebook development, theme refinement, and interpretation of de-identified excerpts.
Workshops 2 and 3 were coded independently by the two joint first authors, followed by collaborative sessions where all three authors compared interpretations, discussed discrepancies, and revised the codebook.
Calibration meetings occurred weekly during data collection and continued for three months afterward. 
Disagreements were resolved through constant comparison and privileging participants' phrasing. 
Our inductive coding approach enabled us to stay close to participants' language and actions; we generated 1,051 initial codes across workshops and interviews. 
We did not compute inter-rater reliability, consistent with approaches emphasizing the interpretive nature of qualitative coding~\cite{mcdonald2019reliability}.

Next, we iteratively developed 23 categories capturing recurring patterns (e.g., refusal of AI, concerns about AI use, policy-based concerns) and subcategories (e.g., gray areas, citing requirements, constraints on creativity). 
To surface relationships, the research team constructed an affinity diagram in Miro~\cite{Miro}, with cluster boundaries negotiated in group meetings. 
Throughout, we maintained an analytic memo corpus (19,408 words) reviewed in weekly meetings to stabilize interpretations.
We judged categories conceptually sufficient within this study's bounded context---a single graduate design studio course---when later data elaborated existing categories rather than generating new ones; we treat our analysis as offering a coherent account of this cohort's experiences rather than claiming generalizability to all students or institutions.
Finally, we remained reflexive about positionality. 
The two student co-authors contributed insider perspectives as peers, while the faculty PI remained distanced from identifiable transcripts to reduce the influence of faculty authority. 
This insider–outsider pairing supported proximity to student voice while preserving critical distance.

\begin{figure*}[t]
  \centering
  \includegraphics[width=0.38\textwidth]{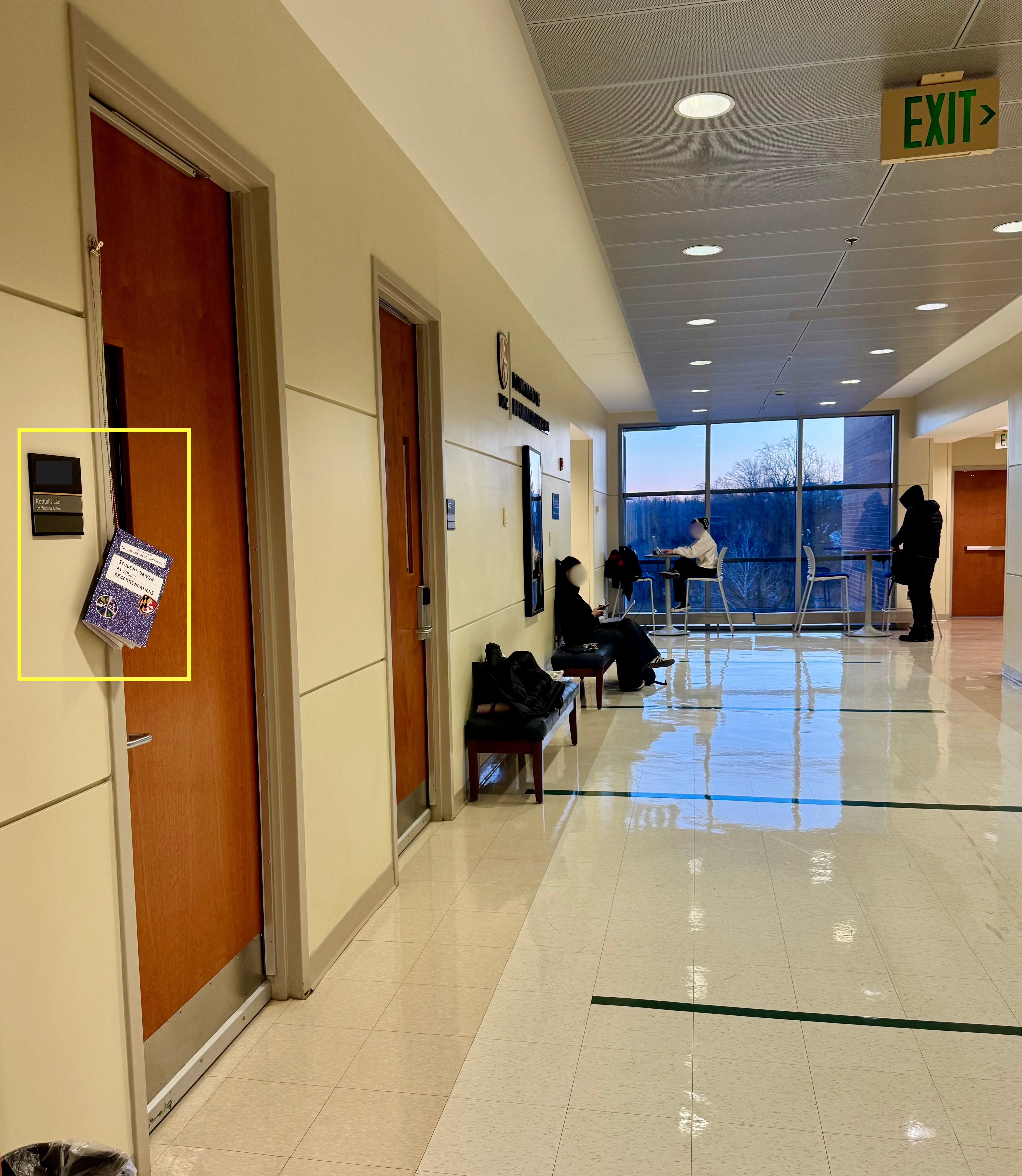}
  \hspace{0.7cm}
  \includegraphics[width=0.38\textwidth]{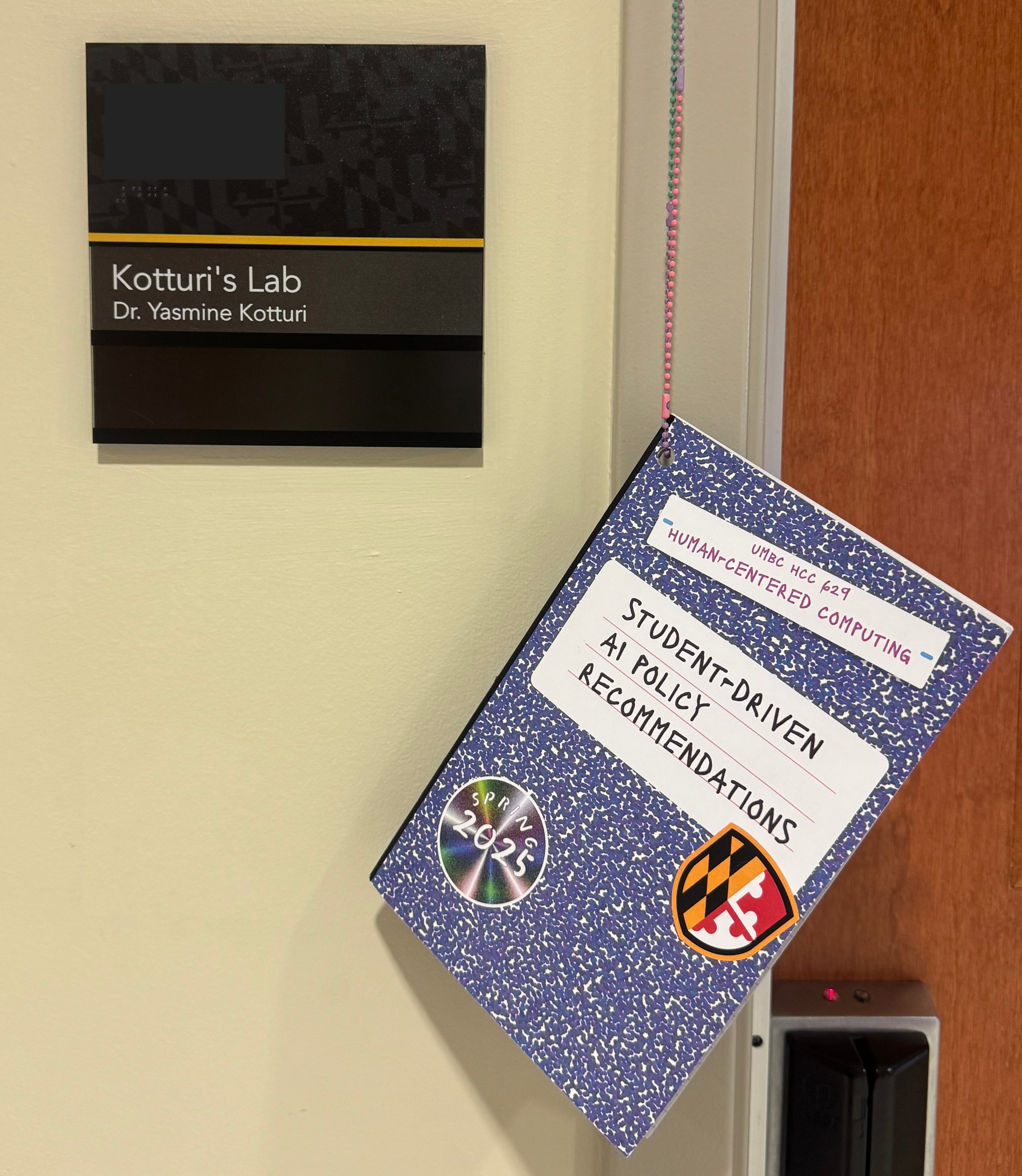}
  \caption{The Student-Driven AI Policy Recommendation zine displayed outside of the research team's lab (right), in the main hallway of the building (left). This central location facilitated student discussions beyond the bounds of the workshops. In addition, over 300 copies of the zine were printed and dispersed throughout UMBC's campus.}
  \label{fig:zines}
  \Description{Two photos showing the Student-Driven AI Policy Recommendations zine displayed in an academic building hallway. Left: A wide view of the fourth-floor hallway showing the zine hanging from a door marked "472 Kotturi's Lab, Dr. Yasmine Kotturi," with students in the background studying near windows. The zine's location is highlighted with a yellow rectangle. Right: A close-up of the zine, which resembles a composition notebook with a purple speckled cover. The zine features handwritten-style labels reading "UMBC HCC 629 Human-Centered Computing" and "Student-Driven AI Policy Recommendations," along with a "Spring 2025" holographic sticker and the University of Maryland, Baltimore County crest.}
\end{figure*}
\section{Findings}
We structure this section around three key findings: 1) how the workshops supported candid disclosure of AI practices and rationale, grounded policy discussions in recent lived experiences, and scaffolded policy articulation resulting in, 2) ten student-authored policy recommendations, and finally, 3) how both the policies and process of participation sparked reflection among participants---and the campus community more broadly---towards more intentional AI use and the importance of participatory governance of AI.

\subsection{Setting the Stage for Policy Making: Candid Disclosure of AI Use among Students}
The workshops created conditions for candid peer dialogue that participants described as uncommon in their everyday academic experiences (P1, P2, P6, P8). As P6 noted: 
\begin{quote}
    ``It's not common to talk about what we're all using and doing with AI. This is a nice opportunity to hear those perspectives.''
\end{quote}

P8 emphasized that despite AI's controversial status, the workshop provided space to \textit{``discuss openly and freely''}: \textit{``I think AI is controversial for some part, but it's very essential these days, many companies and many students and many people are using it. So I think it is a great opportunity to discuss openly and freely.''}

Notably, participants were often unaware of how their peers used AI until the workshops surfaced these practices. 
When a member from the university's AI committee presented campus-wide survey data, participants expressed surprise at the breadth of adoption. 
P1 shared, \textit{``I find it surprising that even visual arts students are using genAI quite a lot. I didn't really expect that.''} 
P2 followed up with \textit{``The psychology students are using the GenAI. That's a good thing and a bad thing.''}—a comment that sparked reflection on AI adoption in fields like psychology, where students might go on to become therapists; for P2, this realization was unsettling.

While faculty concerns regarding student overreliance are well-charted in the related literature, thick descriptions of students' concerns towards overreliance of AI systems are less understood. 
Most participants described their use to overrelying on AI and expressed concerns about this dependency [P1, P2, P5-P8].
At the crux of this overreliance was the pressure of deadlines and feeling overextended. 
Due to the time pressure or their self-prescribed laziness, they ended up using AI.
Sometimes, participants considered AI as a safety net, leading them to habitual cognitive offloading and hindering independent idea and content generation. 
For instance, P7 shared: \textit{``We are busy in those cases wherein we need content ideas, brainstorming. Our mind has gone into lazy mode and it has become a habit.''}
P8, who used AI for reading summarization, expressed concern because they relied on AI to read papers rather than their own reading comprehension; they felt guilty delegating work to AI that instructors expected them to do.
Participants shared various concerns about using AI as it related to their mastery of the course's design material. 
For instance, regarding their comprehension of concepts, their concerns echoed the faculty and administrators who typically dictate policy: they warned that certain uses---generating responses entirely [P1, P6], direct answering [P3], unexamined edits [P2], and summarization [P8]---displace the thinking that leads to conceptual grasp.

\subsubsection{Hidden Practices: Policy Violations and Gray Areas}
There were instances where participants reflected on how and why they used AI in ways that they knew clearly violated their course's AI policy [P1, P3, P5, P7]. 
Some described relying almost entirely on ChatGPT (e.g., ``80-90\%'', P7) to complete assignments with little personal contribution, while others shared how they incorporated AI-generated outputs without citation despite clear prohibitions. 
Participants concealed their usage because they either knew it was unauthorized, were unsure if it was allowed, or felt guilty about being unable to complete the assignment independently. 
One participant noted that even in an assignment designed to critique AI's perspective, classmates turned to ChatGPT to write their critiques [P3]. 
P5 described a specific instance where they used AI in a way that was disallowed:

\begin{quote}
``I used [AI] to write a discussion [post]. And that day...I just wasn't feeling it, and it didn't involve any sources at all. I just said [to the AI tool], `Answer this for me.' I just pasted it. And I know some professors have AI detectors, so they probably just put it through there. I was like, `Yeah, you know what, I'll take the L that day.' So that's what happened...they just put in the comments, `I could just tell this wasn't you...' Instead of 100[\%], they gave me an 80[\%].'' 
\end{quote}

While this incident did not occur in \coursename, it points to an interesting phenomenon: P5 received an 80\% on an assignment that was entirely AI-generated and in clear violation of their course's policy. 
This mismatch between the rules and application of penalties may be contributing to the confusion and disarray, and warrants specific attention, as we further explore in the discussion section.

\subsection{Ten Student-Driven AI Policy Recommendations} 
Facilitating candid conversations allowed participants to brainstorm, apply, and refine policy recommendations. 
In this section, we define each of the ten policies and provide details such as disagreements among participants or challenges experienced when participants put their policies into practice.

\subsubsection{\textbf{Policy Recommendation \#1 Instructions:} Instructors should include guidelines with concrete examples of acceptable and not acceptable use of AI for each assignment.} See Figure~\ref{fig:zine_instruc} in the Appendix for corresponding zine page.
\label{sec:instructions}

Participants felt frustrated by the lack of clear guidelines on acceptable AI use. 
The AI policy in \coursename allowed students ``to use AI lightly'' but required citation via chat logs with ``strict enforcement'', mirroring a common vagueness found in existing AI policies~\cite{smit2025ambiguous,tsao2025trajectories}.
Students were uncertain what this meant in practice. 
After re-reading this policy, P3 still wondered: \textit{``how much can we use AI? What does `light use' mean?''} 
As P1 and P2 emphasized, sometimes their AI tool would alter or generate too much, taking their use into a gray area without their approval, making it hard to ``undo.''

This ambiguity worsened when enforcement did not align with the stated policy. 
As noted earlier, P5 received 80\% for a submission entirely generated by AI in clear violation of course policy. 
Similarly, P10 reflected on peers who used AI to create video submissions despite explicit prohibitions, yet faced no consequences: \textit{``...So what [did] we learn, that [students] can use AI in the video, even though they were told not to use it, and they can still pass the class. Next semester, if someone says, `Hey, don't use AI for this assignment, otherwise you won't pass the class.' And if they still use AI for the assignment, they're expecting to pass that class.''} 
This disconnect between policy and enforcement made AI guidelines even more tenuous.

To address this ambiguity, P1 suggested providing concrete examples:
\begin{quote}
     ``[Instructors] can provide acceptable and unacceptable use of AI...in the starting of the class. Run through a presentation and give [students] acceptable and unacceptable AI examples.'' 
\end{quote}

P1 noted this could also strengthen esprit de corps: \textit{``I feel like it should be done because it makes life easier. If you leave students on their own, they probably will be confused.''} 
P6 agreed that examples would reduce fear around citation: \textit{``If someone is using a lot of AI and they fear not citing it, in order to remove that fear, like [P1] said, a brief example would do it.''}

Finally, participants noted that blanket course-level policies often lacked the granularity needed across assignment types. 
Tool-specific policies—such as banning ChatGPT but allowing Grammarly—were also becoming outdated as more platforms quietly integrate generative AI features, reinforcing the need for assignment-level guidance.

\subsubsection{\textbf{Policy Recommendation \#2 Ownership}: We should only use AI for up to 50\% of our work on any given assignment, so that the majority reflects our own ideas and effort.} See Figure~\ref{fig:zine_owner} in the Appendix for corresponding zine page.
\label{sec:ownership}

Ownership---also referred to as ``authorship'' and ``accountability'' throughout the workshops---generated rich, sometimes contentious discussion among participants. 
Students naturally framed ownership in terms of percentage: how much could be AI-generated while still being considered their work? 
There was disagreement on this threshold, captured in the corresponding zine page on ``post-it notes''.
P4 argued that at most 30\% should be AI-generated; P1, P6, and P9 proposed 50\%, concerned that higher thresholds would be exploited; P7 and P3 argued the percentage was irrelevant as long as the original ideas came from the student. 
Ultimately, 50\% emerged as a consensus halfway point.

However, after applying this policy in the final workshop activity, P3 shared how this policy was too rigid in their post-workshop reflection:
\begin{quote}
``Honestly, the biggest issue is that [the ownership policy] feels a bit rigid, especially when you're not using AI to do the thinking for you, but just to move faster. Like in my case, I had a clear mental model, and the AI tool just helped me skip the dragging-and-dropping part. But if we follow the current rules strictly, that might still count as `more than half AI-generated,' even though the actual decisions came from me. That's where it gets tricky.''
\end{quote}

P3 disambiguated different kinds of use, noting that not all AI use should be policed the same way—distinguishing use that helps ``move faster'' when one already has a clear vision from use where AI acts as a collaborator (a concept described in detail in~\cite{Satyanarayan2024Intelligence}): \textit{``...I wasn't blindly accepting what AI gave me. I already had the structure in mind, and AI just helped speed it up. It felt like streamlining, not outsourcing.''}

P1 noted that as heavy AI use becomes normalized in industry, such thresholds may need reevaluation to support employability. 
P6 pointed to the labor required to maintain ownership when AI suggestions were strong: \textit{``I had to consciously revise and rewrite AI-generated text to make sure it reflected my own thinking. Balancing usefulness with ownership required effort, especially when the AI suggestions were strong.''}
While participants were open to following this policy, they admitted compliance would depend on context.

Finally, another challenge was measurement: how can someone calculate the percentage of AI use across multiple phases of the design process?
Relatedly, and connecting this issue to Policy \#1 on Instructions, P4 noted that examples would clarify what 50\% actually looks like in practice: \textit{``This policy needs to be explained to all students with this image generation and text generation example to show what less than 50\% actually means, otherwise, it didn't even make sense to me.''}
So while all participants agreed that ownership over their own work was essential, the particulars of how to conceptualize and quantify this proved challenging; we revisit and delve into this further in the discussion section. 

\subsubsection{\textbf{Policy Recommendation \#3 Divergent Thinking}: When using AI for brainstorming, we should push ourselves to explore alternatives, surprising directions, or ideas that feel more personal and meaningful to us.} See Figure~\ref{fig:zine_diverg} in the Appendix for corresponding zine page.
\label{sec:divergent-thinking}

Participants gravitated toward divergent thinking—the essential stage in design that focuses on generating many ideas before converging on a subset to pursue~\cite{Tohidi2006}. 
Participants noted that AI use could both hinder self-expression and creativity [P1, P3, P4, P5, P7, P8] and support it [P2, P6], depending on context—especially when deadlines or workload would otherwise block deeper engagement.

When applying this policy during the design activity, several participants reflected on using AI to support ideation. 
P1, P2, and P5 noted that the policy encouraged them to prompt AI to branch ideas, explore alternatives, and resist shallow responses. 
P2 shared:  
\begin{quote}
    ``I used GenAI primarily as a creative amplifier, especially in the ideation phase...My approach remained iterative. I would prompt the model, reflect on its output, and then reframe or refine ideas to suit my intended direction. The back-and-forth helped me stretch my thinking without losing authorship over the outcome.'' 
\end{quote}

P8 felt AI could theoretically help when peers are unavailable for feedback, but found the outputs too generic in practice: \textit{``If I don't have any peers to brainstorm together, [then] I could ask [AI]. [But] I feel like it gave me too broad, too general viewpoints.''} 

P10 offered a dissenting perspective, questioning whether AI-assisted design undermines skill development:
\begin{quote}
``As long as you have the fundamentals down—you should at least have those down—if you're using AI for every step of the design process, are you doing the work? Because you can say that, `based on my personal taste, I like to do this,' but then are you doing the work? Or is AI doing the work at that point?''     
\end{quote}

P10's concern points to a consistent challenge of overreliance on AI and lack of skill building: AI can prevent students from developing the foundational skills they need to do the work independently.

\subsubsection{\textbf{Policy Recommendation \#4 Job Skills}: We should be provided opportunities to learn how to use AI in ways that reflect how AI is used in real workplaces through coursework. Instructors should stay updated on how AI is used (and regulated) in industry.} See Figure~\ref{fig:zine_job} in the Appendix for corresponding zine page.
\label{sec:job-skills}

Participants noted that most students learned to use AI on their own rather than through formal instruction, resulting in significant knowledge gaps: P7 was concerned about identifying AI-generated bias; P1 felt unprepared to write effective prompts; P3 noted that lack of training contributed to academic integrity concerns. 
Currently at UMBC, as at many institutions, AI training is not provided to students.

Participants argued that any training should center employability. 
P1 shared how their roommate's new employer expected all employees to use AI from ``day one''. 
P6 reflected that AI skills were now required for UX job applications—the career path most participants intended to pursue. 
P7 added that AI literacy was also important for research careers and Ph.D. programs.
Alongside these needs, participants grappled with the lack of support their university and instructors provided.

However, P5 raised an important caveat: in some jobs, such as government positions, AI use is strictly prohibited. 
Given the university's geographic proximity to government employers, this constraint was relevant for many alumni: \textit{``...what's the point of teaching it if we can't even use it in the first place?''}. 
P5 emphasized that any training or guidance should account for these job-specific considerations.

P10 pushed back on this policy entirely, arguing that the course was meant to teach fundamental design skills, which should not be supplemented with AI assistance; students should be able to complete tasks independently in case AI becomes unavailable—such as when transitioning to a job where AI use is prohibited, as P5 described.

\subsubsection{\textbf{Policy Recommendation \#5 Bias}: We should check if AI-generated ideas include any stereotypes or biased assumptions, such as by asking if any perspective or voice is missing in the response received.} See Figure~\ref{fig:zine_bias} in the Appendix for corresponding zine page.
\label{sec:bias}

``Bias'' was one of the policy topics participants did not initially gravitate toward—not due to lack of interest, but lack of practical support. 
When reflecting on this policy in practice, P1 struggled with the labor required, noting no commercial tools exist to help: \textit{``I had to carefully go through the content, analyze language and representation, and possibly cross-check against standards or guidelines of the work to ensure that no biased assumptions or stereotypes were present, which was time-consuming.''}
Other participants found the policy easier to follow. 
P2 shared: \textit{``I stayed away from anything that felt biased or vague—if something felt off or too generic, I just changed it.''} 
P4 took a different approach, relying on AI itself to check for bias: \textit{``I always ask AI if there is any bias and it is usually quick to detect that, so that's why I like adhering to this policy.''}
While participants differed in their strategies for addressing bias, their reflections point to a broader need for accessible tools that help students identify and critically assess biased content in AI outputs.

\subsubsection{\textbf{Policy Recommendation \#6 Citing AI Use (or Not)}: It is important for us to credit AI where credit is due. However, sharing chat logs, as many instructors currently require, is tedious and ineffective for both students and graders. Instead, we should share a 2-3 sentence summary for each submission describing how and why we used AI} See Figure~\ref{fig:zine_citing} in the Appendix for corresponding zine page.
\label{sec:citing}

Citing AI use was one of the most debated topics across workshops and post-workshop reflections. 
Almost all participants recognized that citing via chat logs was tedious; P5 called it ``a big ask,'' and P7 believed AI use should not require citation at all.

P9 offered a counterpoint, arguing that chat logs—even partial ones—would make dishonesty harder and give graders a clearer sense of how much AI was used. 
While acknowledging that instructors may not read every log in detail, P9 felt that even skimming would reveal red flags and considered logs more transparent than brief written summaries.

P6 highlighted a potential benefit of citation: if students cited their AI tools, others could learn what tools and approaches their peers were using. 
As P6 noted: \textit{``I think if they get a hold of it, or if they see someone else using something different from them. That's where they realize, if they cite it, it's gonna be useful for them and others too, when they collaborate and when they realize the benefits of it.''}

However, most participants resisted citation requirements, pointing to disconnects between policy and intention. 
P4 shared how they circumvented the requirement by prompting AI to cite itself:
\begin{quote}
``One of the policies is basically, `[cite] everything everywhere we use it.' Basically, if you tell GenAI `Consider the policy and then cite yourself appropriately wherever you are.' It does that whether it's citing or paraphrasing or putting the references that will cite itself.''
\end{quote}

P5 and P7 raised a deeper concern: citing AI use meant, for them, revealing how little of the assignment was completed independently. P5 asked:
\begin{quote}
``Is there such a thing as being too transparent? Because if the professor is asking [you] to cite the use of AI, and you have AI do the whole project for you, and you cite that, then you're cooked...But versus, if you use AI to edit a line of your paragraph, or a line of code, and you still cite it, you get points taken off.''
\end{quote}

Tucked behind what appears to be overreliance on AI, P5 is also pointing to the uneven enforcement of AI restrictions, within and across their courses. Given this unevenness, it further pushed them towards more opaque use of AI (the logic being: if I am going to get dinged for any AI use, big or small, then there is no incentive to report truthfully. P7 candidly chimed in on a related note:

\begin{quote}
``So one major problem I had in \coursename  [was that] sometimes I used 80 to 90\% of ChatGPT to do the assignments, [and] only my [completed portion was] 10\%. So I had a fear, [if I have to cite], I cannot cite everything, right? It will look like I'm just using ChatGPT for the assignment. It shows that you have done nothing, right?''
\end{quote}

Both P5 and P7's candid reflections highlight a key motivation for understanding why students do not include chat logs and report AI usage, even when more experimental policies are in place. Given this ineffective citation mechanism, P1, P2, and P6 proposed an alternative solution which may prompt more reflection and critical use, while also making adherence to citation policies more likely: include a short summary describing how and why AI was used.
To capture the range of views, the corresponding zine page included post-it notes with dissenting perspectives (see Figure~\ref{fig:zine_citing} in the Appendix, page~\pageref{fig:zine_citing}).

\subsubsection{\textbf{Policy Recommendation \#7 Hypocrisy in Faculty Use}: If instructors expect us to be transparent about our use of generative AI, we expect the same transparency from them when it comes to how they use AI in teaching, grading, or creating assignments.} See Figure~\ref{fig:zine_hypoc} in the Appendix for corresponding zine page.
\label{sec:hypocrisy-faculty}

Students perceived a double standard: their AI use is highly regulated, while those who dictate policy face no similar requirements. 
P4 argued that transparency should apply equally to students and faculty, adhering to the golden rule. 
Beyond equitable treatment, students wanted to understand how instructors' approaches might be changing in response to AI—and what that meant for their learning. 
One student leader reflected that if faculty use AI to free up time on rote tasks, that time should be reinvested in one-on-one instruction.

P9 took a stronger stance, arguing that faculty should demonstrate expertise by producing teaching materials themselves rather than ``cheating'' with AI:
\begin{quote}
``If you're using [AI] to teach, get your Ph.D. revoked, honestly. You should know how to teach the content that you wrote a dissertation on. I don't think that would be in any way acceptable.''    
\end{quote}

While P9 was fine with minor uses like grammar or formatting, they emphasized that students pay tuition for professors' knowledge, not AI's output.

\subsubsection{\textbf{Policy Recommendation \#8 English Learners}: For those of us who are English learners, we should be encouraged to use AI to support our English proficiency in writing by asking for refinement of our text. Also, we should ask for an explanation of the refinements made to further our language proficiency.} See Figure~\ref{fig:zine_english} in the Appendix for corresponding zine page.
\label{sec:english-learners}

P1, P2, P7, and P8 expressed difficulties conveying their thoughts in English, given that it was not their primary language. 
With limited vocabulary and grammatical structures, their writing often felt overly simple, and they expressed frustration at being unable to articulate ideas as clearly as they wished. 
For these participants, AI tools were almost necessary to support writing at an academic level.

Participants emphasized that policies should explicitly recognize AI's role in helping English learners understand context and fully express ideas—including emotional nuance—without penalty. 
P2 noted that the level of fluency should be factored into any policy, a point echoed by one of the student leads, also an English learner. 
P5 and P8 observed that even native English speakers use AI for grammar and spelling support, suggesting the line between acceptable and unacceptable use is blurry for everyone.

However, P9 cautioned against treating AI as a language learning tool. 
They acknowledged that AI could help when students struggle to put thoughts into English, but warned that English learners might accept AI suggestions at face value—even when the wording or tone is inaccurate—inadvertently learning mistakes. 
As P9 put it: \textit{``You don't know the words to say, and AI is giving you something. How do you know AI is giving you the actual words you want to say?''}

Ultimately, this policy aimed to ensure that students less proficient in English are not penalized for leveraging AI, while recognizing the need for critical evaluation of AI suggestions.

\subsubsection{\textbf{Policy Recommendation \#9 Feedback}: We can use AI to help us better understand and respond to peers' feedback, especially when revising our work.} See Figure~\ref{fig:zine_feedback} in the Appendix for corresponding zine page.
\label{sec:feedback}

This recommendation received unanimous agreement during the initial policy review.
By reframing AI as a feedback intermediary, the policy positioned AI not as a replacement for peer insight, but as an aid in making feedback more digestible and growth-oriented—especially in classrooms where social dynamics inhibit candid exchange.
Because participants did not exchange feedback on each other's design assignments in Workshop 3, they were unable to put this policy into practice, resulting in less discussion to capture.
One emergent challenge during the making of the zine was how to visually represent this policy. 
Students proposed depicting AI as a ``shield'' for feedback—initially imagined as a Captain America–style shield deflecting harsh comments, then refined into a filter that softens harmful feedback while translating vague critiques into actionable guidance. 
This metaphor framed AI as a mediator that scaffolds emotionally intelligent exchanges, addressing concerns raised throughout the workshops: discomfort with critique, unclear peer review norms, and the emotional labor of giving and receiving feedback.

\subsubsection{\textbf{Policy Recommendation \#10 Equity}: Everyone in class should have access to the same AI tools or models for each assignment to ensure fairness.} See Figure~\ref{fig:zine_equity} in the Appendix for corresponding zine page.
\label{sec:equity}

This policy addressed potential inequities in access, as some students can afford more powerful AI tiers than others. 
Interestingly, the original version of this policy was more prescriptive: ``the same AI tool and model should be used by all students for each assignment.'' 
However, upon applying the policies in the third workshop, participants noted loopholes and found it too restrictive.
P4 shared: if a policy mandates identical tools, students who build their own AI tools—demonstrating technical creativity—would technically be in violation.
P7 emphasized that standardizing access—providing a baseline set of tools to all students—would ensure fairness and help students navigate the rapidly expanding AI landscape. 
However, P3 noted the challenge of maintaining such a list: \textit{``There are a lot of AI tools right now. People don't even know the existence of them.''}
P10 questioned whether students would receive institutional AI subscriptions and what ethical standards should guide that selection. 
P9, who does not use AI tools, argued it would be unfair to pay for a subscription they would not use—similar to the gym fee the university already charges them.

\subsection{From Policy to Practice: Shifts Toward Intentional Use}
Participants shared that the zine-making process created a sense of permission to question and debate classroom AI policies. 
One of the most notable shifts after the workshop series was participants' increased intentionality in how they used AI.
Many described changes in their practices, often referencing the conceptual metaphors from the design activity—such as AI could be used as a reflective or critical design partner [P1, P7], a feedback generator [P1], or a collaborator [P1, P2, P6].

Strategies for intentional use included shifting from passive to reflective collaboration [P1], bringing more critical insight to their work [P2], and prompting AI iteratively throughout a task rather than with a single prompt [P6].
P2 shared: \textit{``This time, I used [AI] more intentionally and strategically. I treated GenAI as a collaborative thought partner to question my assumptions, reframe ideas, and push past obvious solutions.''} 
P6 echoed this: \textit{``GenAI served as a thinking partner helping to speed up decision-making without doing the thinking for me.''}

Rather than using AI simply to complete tasks, participants worked with it collaboratively—approaching use with intentionality.
P7 suggested that institution-wide AI policies should serve not just to penalize misuse, but to increase awareness and intentionality:
\begin{quote}
``I learned it is very important to have AI policies in the institute, because students are not aware how they're using AI, how much they want to use, and how they can use Gen AI [tools]. Having such policies and having such guidance and instructions will help students to be aware of the future opportunities, future challenges and risks.'' 
\end{quote}
Involving students in the process of creating policy allowed them to reimagine how AI could support—rather than undermine—their learning, a point which we revisit in the discussion section.

\subsubsection{Ripple Effects across the Campus}
The zine, policy recommendations, and ethos of centering student perspectives circulated beyond the workshop context, catalyzing continued conversation among faculty, students, and senior administrators.
Faculty across disciplines drew on the zine as a prompt for classroom practice. 
Two HCI faculty guided discussions on AI policies with their students; a computer science professor brought the zine as a starting point for co-designing AI policy during the first week of an AI ethics course.
Other faculty encountered the zine hanging outside the research team's lab (See Figure~\ref{fig:zines}).
The project also seeded a new research collaborations: one participant began working with faculty on a related project, and one student leader is now collaborating with the university library to run a campus-wide series of these workshops across disciplines, extending the participatory model beyond a single domain and course.
The zine has since been archived in UMBC's Library---both physical and digital versions\footnote{\url{https://lib.guides.umbc.edu/c.php?g=1475372&p=11187429}}---ensuring that student perspectives persist in institutional memory beyond the workshop series.

However, these ripple effects also revealed persistent misunderstandings of the project's goals.
When the project was presented to two faculty members in the research team's department, they initially interpreted the zine as a mechanism for increasing compliance---helping students adhere to rules and improve grades.
We stress that the goal of this project is not strictly enforcement-oriented but rather centering students as experts, or ``lead users,'' within this context~\cite{von2006democratizing}.
These misreadings underscore how deeply ingrained top-down framings of AI policy remain, even among faculty sympathetic to student perspectives.
By tracing these follow-on effects, we see how the zine extended workshop conversations into everyday pedagogical practice—while also surfacing the ongoing work required to shift institutional norms.

\section{Discussion}
\change{
In this discussion, we re-situate the student-driven policy recommendations in contemporary discourse around university AI policy---a rapidly evolving space~\cite{kuoPolicyCraftSupportingCollaborative2025, puHowCanWe2025}. 
In particular, we consider how these student recommendations fit into the existing spectrum from experimental policies---where AI is integrated into classrooms as a supportive tool---to restrictive policies---where AI is viewed primarily as a threat to traditional learning methods~\cite{sumilong2025instructional}.
To do so, we first unpack what practical challenges exist when implementing students' policies (such as when policies might not be synergistic with learning goals).
Then, we consider what the value of student-driven recommendations may be (while remaining cognizant of the aforementioned challenges), such as how the process of participation, rather than solely the resulting policies, is an essential and transferable component of this work.
Finally, we conclude with recommendations for how to engage students' perspectives on AI policy in higher education.}

\change{
\subsection{Challenges with Implementing Student-Driven Policies}
In this section, we discuss challenges with implementing student-driven policies, beginning with challenges related to feasibility, including enforcement, overhead, and lack of data. Then, we discussed challenges that arise when there may be an incongruence between students' policies and course learning goals. 
}

\change{
\subsubsection{Feasibility challenges with student-driven policies}
We consider several of the students' policies---``Equity'', ``Ownership'', ``Job Skills'', and ``Instructions''---which may not be feasible to implement for various reasons such as enforcement difficulties, technical limitations, large overhead for instructors, or simply, lack of know-how.
To begin, with Policy \#10 on ``Equity''  (See Section ~\ref{sec:equity}), participants sought equitable tool access across their peers, so that no one student had a particular advantage by accessing expensive or lesser-known AI tools. 
In some ways, restrictive AI policies address this issue: by disallowing all use of AI, students do not have uneven access to AI tools~\cite{de2024generative}.
Instructors with experimental policies, on the other hand, have attempted to tackle this issue by co-creating statements with students at the beginning of their course, where all students agree upon which tools they will and will not use ~\cite{uq2023educ3707syllabus}.
}

\change{
Ultimately, one of the practical challenges across all of these approaches towards equity is enforcement ~\cite{gonsalves2025addressing}: how to make sure students do not use the tools that their policies, either co-created or not, disallow them from using.  
In order to understand how to address issues of enforcement, we argue it is essential to dig into the reasons why students do not abide by policies in the first place. 
In the case of equity, one participant shared that they created their own AI tool to aid with their assignments, and it felt unfair to not allow them to use this, even though other students could not access it. 
By understanding the reasons that students have for not adhering to course policy, instructors can decide an appropriate course of action---in this case, can the student provide an in-class demonstration of the AI tool they built, and even extend use to their peers? Or should the instructor facilitate a discussion for why such personalized tools give that student an unfair advantage?
Not only is such transparency likely to support enforcement, it may also foster more critical thinking and awareness around use of AI tools~\cite{wmu2024ai_critical_thinking, chanComprehensiveAIPolicy2023}.
}

\change{
Policy \#2 ``Ownership'' (See Section~\ref{sec:ownership}) was another policy where students discussed enforcement challenges themselves during workshops: the 50\% threshold, for instance, assumes an AI's contribution can be quantified and distinguished from a student's contribution.
In this case, enforcement relates to a longstanding challenge connected to remixing in design, often summarized in the phrase: \textit{all design is redesign}~\cite{latour2008cautious}.
Participants considered how it is not clear where their ideas stop, and the \textit{AI's} ideas start, especially when rapidly iterating with an AI.
Restrictive policies bypass challenges of student ownership by not allowing any use in order to preserve student ownership~\cite{LibertyUniversityAI}, whereas experimental policies often assume students' responsibility for anything they submit, regardless of what tools have been used~\cite{HarvardProvostAI2023}.
In design pedagogy, understanding the importance of process continues to be a key learning objective~\cite{Tohidi2006}, and process-tracing tools like BoodleBox~\cite{BoodleBox} may provide one path for exploration in future work when attempting to address this policy's implementation challenge.
}

\change{
Another policy with feasibility challenges was Policy \#4 ``Job skills'' (See Section~\ref{sec:job-skills}), as participants wanted instructors to update course materials to include information about what AI skills they need to have in order to land their desired job; often in response to repeatedly witnessing headlines about AI's takeover of knowledge work.
On the one hand, this seems like a simple and reasonable ask.
Increasingly, universities with experimental policies, such as Ohio State's ``AI Fluency'' initiative~\cite{osu2025fluency}, claim connections between their revamped courses, AI literacy, and career readiness.
However, it is unlikely to be that simple:
currently there exists minimal empirical evidence to support claims around AI literacy and AI skills~\cite{reich2025stop}.
In many contexts beyond traditional ML/AI courses, where students learn fundamentals to build, train, and deploy AI models, it is not clear which skills are central and worth pursuing (recall how short-lived the 2023 most in-demand career of prompt engineering was~\cite{bousquette2024hottest}).
In design-related careers, for instance, AI's infiltration is leading experts to both critically evaluate AI-generated outputs, which may be unreliable and misaligned with user needs, and develop effective human-AI collaboration skills ~\cite{li2024user}.
}


\change{
Taken together, even as data becomes available on how AI is reshaping skills and careers, and what AI literacy actually entails~\cite{reich2025stop}, it is then a non-trivial task for instructors to re-imagine and rebuild curricula to incorporate these learnings, especially given the often-lack of institutional support offered to instructors~\cite{barrett2023not, mcdonaldGenerativeArtificialIntelligence2025}.
Similarly, Policy \#1 Instructions (See Section~\ref{sec:instructions}) beckons additional instructor labor.
The overhead to implement assignment-specific instructions for AI use, while ensuring academic integrity, and as AI tools' capabilities are changing rapidly, is a herculean task.
}

\change{
\subsubsection{What Happens when Student-Driven Policies are not Aligned with Learning Goals?}
While scholarship on universities' AI policy is a fast-moving and capricious space, there seems to be one clear signal: students tend to be more lenient than faculty on AI use and more supportive of AI technologies writ large~\cite{10.1002/pra2.1277, sah2025generative}.
One reason for this difference is that instructors are often concerned about students' over-reliance on AI tools, or that AI use will likely impede student learning by cognitive offloading ~\cite{kosmyna2025your}.
In our study, this tension exists in several of the students' policies, such as ``Citing AI use (or Not)'', ``English Learners'', and ``Ownership''.
}

\change{
When it comes to the need to cite the AI tools students use, one relevant learning goal is clear: students need to know how to cite and paraphrase reputable sources~\cite{eragamreddy2025plagiarism, samson2010information}.
Most of the UMBC courses participants had taken required them to submit all chat logs from AI tools as part of their bibliography (similarly to many existing experimental policies~\cite{brown_ai_citation_attribution, cornell_elso_ai_integrity_2025}).
But students noted the tedium, and even hinted at the irrelevance, of this request: not all AI systems they use are embedded in chat-based tools which afford log sharing  (See Policy \#6 ``Citing AI Use (or Not) in ~\ref{sec:citing}).
Instead, students, after lengthy debate, compromised on including a summary of which tools were used and how at the end of the assignment---more or less a process and tools statement~\cite{ConestogaDescribeAI}.
Still, there was strong sentiment that citing AI tools (even in this relatively simple way) is an undesirable and overly tedious task---a stance which is in clear opposition to an essential learning goal of university education.
}

\change{
Policies \#8 ``English Learners'' (See~\ref{sec:english-learners}) and \#2 ``Ownership'' (See~\ref{sec:ownership}) followed a similar pattern where students' policies may diverge from learning goals. 
For instance, some restrictive policies specifically flag that non-native English speakers should not use these AI tools in any way to overcome a language gap~\cite{mcnamara2025socy136}, whereas other universities with more experimental policies, such as at Teachers College Columbia University, state that these tools may present unique opportunities to support this population~\cite{Najarro2023} (which some new startups hope to capitalize on ~\cite{LanguaTalk2021}).
}

\change{
Restrictive policies which touch on ideas of ownership typically disallow use of AI because the very nature of AI tool use complicates students' ability to take responsibility for their work ~\cite{studentaimpact2023policy}.
In our workshops, students' policies supported liberal use of AI among English learners, and complicated notions of ownership to include AI-generated outputs as students' original work---but such ideas may be orthogonal to learning objectives. 
In these cases, it may behoove faculty not to solicit students' perspectives on AI use in order to safeguard their own learning. 
But, we argue, rather than disallowing use sans explanation, it is critical to provide an accompanying rationale for such restrictions in order to encourage students to stay within the bounds of acceptable use.
}

\change{
\subsection{The Benefits of Student-Driven Policies}
In this section, we detail what the benefits of student-driven policies may be, while remaining cognizant of the challenges described above.
We argue there are two key beneficial aspects to student-driven policies: 1) the new perspectives students contributed by reflecting on their lived experiences as early adopters of generative AI technologies, and 2) that the process of participation itself is a valuable step towards a future of AI governance in higher education that serves all stakeholders. 
For the latter, we provide actionable suggestions for how to scaffold such participation.
}

\change{
\subsubsection{Students Contribute New Policy Ideas and New Perspectives on Existing Policies}
During workshops, student-led discussions enabled fresh takes on existing policy topics as well as new policy topics that may be important for policy designers to consider.
Take Policy \#9 on ``Feedback'' for example (See Section~\ref{sec:feedback}): while some experimental AI policies state how AI may be used by instructors to provide feedback to students on their assignments~\cite{UCDavisPAIRR2026}, students in our workshops discussed how they could use AI to digest instructor or peer feedback and translate it into actionable suggestions or to make it less harsh (an issue with peer feedback in particular~\cite{hicks2016framing, kulkarni2015peerstudio}).
}

\change{
In addition to fresh perspectives on existing policies, both student participants and student leaders---while in conversation with workshop participants---captured policy topics not currently included in university policies.
For instance, while both experimental and restrictive policies emphasize university-wide or course-wide AI policies~\cite{an2025investigating}, participants emphasized how policies needed to also be \textit{assignment-specific} and provide examples of what acceptable AI use is; captured in Policy \#1 Instructions (See Section~\ref{sec:instructions}). 
Participants sought more granular information about what is and is not allowed to assuage anxieties around academic integrity accusations, especially given how participants pointed out how some existing policies are overly vague ~\cite{gonsalves2025addressing}.}

\change{
Poignantly, participants discussed at length their frustrations about existing AI policies, especially restrictive policies that ban AI use among students, but where instructors are permitted to use AI ~\cite{Hill2025NYT}. 
In other words, participants discussed a double standard (what student leaders referred to as ``hypocrisy in faculty use [of AI]'' (See Section~\ref{sec:hypocrisy-faculty})---yet another policy consideration which is often absent in university AI governance discourse, both restrictive and experimental ~\cite{Hill2025NYT, kumar2023faculty}.
Increasing the diversity of stakeholders who design and govern the systems that impact them leads to more effective design and governance outcomes~\cite{muller2007participatory, harrington2019deconstructing}.
In our work, we investigated what this could like in AI governance in higher education, where students' perspectives are not included in the decisions that govern them~\cite{bowenTeachingAI2024,puHowCanWe2025}.
By engaging student perspectives and empowering an overlooked stakeholder group (albeit within the context of one course at one university) led to new policy ideas and fresh perspectives on existing policies.
}

\change{
\subsubsection{Additional Benefits of Student Participation beyond Resulting Policies}  
Thus far, we have focused on the products of student participation---the policies themselves---and their associated benefits and challenges. 
We now turn to the process of participation, and argue that engaging students in AI governance carries value beyond the resulting recommendations such as improved student engagement, critical thinking, and compliance, improved student-instructor relationship building (in a moment where trust is fraying bi-directionally~\cite{gorichanaz2023accused, luo2025does}) and even the opportunity for impact beyond the bounds of the workshops by shifting students from passive policy subjects to reflexive public's capable of ongoing intervention in institutional AI governance~\cite{dantecInfrastructuringFormationPublics2013}.
}



\change{
The ``Students-as-Partners'' (SaP) paradigm shows that engaging students as active collaborators in teaching and learning practices comes with a host of additional benefits beyond improved efficacy such as enhanced perspective taking and increased academic motivation~\cite{matthews2016students, cook2014engaging}.
This is because co-creation shifts classroom dynamics from hierarchical to collaborative, and calls on students to triangulate and reflect on the status quo, their lived experiences, and imaginatively brainstorm improved futures ~\cite{cook2014engaging}.
In our study, engaging students-as-partners facilitated students' understanding of the tensions that exist within AI policy making decisions, as well as the reasoning behind existing policies.
In addition, engaging students-as-partners facilitated our understanding as researchers about the kinds of subtle barriers that may exist when fostering student participation in AI governance: recall participants' hesitation to ask clarifying questions about their instructors' AI policies, as they feared this could be interpreted as an admission of misconduct: \textit{questions are confessions}.
}

\change{
While the scope of our study is limited to one graduate design course at one university, these initial explorations point to an exciting potential of students-as-partners in AI governance in order to facilitate their engagement, critical thinking, and even compliance; when a student understands the reasoning behind a rule, they are much more likely to comply~\cite{steingut2017effect}.
In addition, students-as-partners in AI governance may also have the indirect benefit of improving relationships between students and instructors~\cite{bovill2011cocreators, mercermapstone2017systematic}. 
Given existing breakdowns of bi-directional trust among students and instructors~\cite{puHowCanWe2025}, participatory approaches to AI governance may help to repair breakdowns by calling students \emph{in} rather than calling students \emph{out}.}

\change{
We can also understand students' participation in the context the formation of \emph{publics}, or, in other words, how zine-making and zine-sharing catalyzed participation which persisted beyond the bounds of the workshop~\cite{dantecInfrastructuringFormationPublics2013}.
In our study, zines created conditions for such a public to take shape, comprising not only students but those concerned about the lack of student voice in AI governance: recall the several instructors who, after flipping through the zine, began to explore ways to solicit student voice in AI policy.
Zine-making in particular, was an important artifact for this public to take shape because institutional feedback mechanisms---surveys, comment forms, office hours---often function to absorb concerns without addressing them~\cite{ahmed2021complaint}; evidenced by the plethora of large-scale surveys soliciting student perspectives on AI ~\cite{TytonPartners_TimeForClass2025}, and very little evidence that these data are translated into governance changes ~\cite{mcdonaldGenerativeArtificialIntelligence2025}.
Instead, our zine-based approach sought to create conditions where students could articulate discontent and concerns outside these traditional channels.
}

\change{
\subsubsection{Recommendations for Fostering Participatory AI Governance in a Classroom near You}
We understand that zine-making is not always feasible or desirable.
Other approaches to foster student participation include campus-, department-, and course-level interventions.
For larger-scale participation projects, interventions such as listening sessions or town halls may be ideal, especially when interdisciplinary discourse on AI policy may be desired.
For course-level interventions, instructors could begin class with a co-articulation of policies and agreements, and then solicit students' formative feedback midway through the term on how AI policies are and are not working (and why).
Policies can be embedded into other artifacts, besides zines, better suited to different disciplines: one-pagers for professional programs, podcast episodes for communications students, GitHub repositories for computer science students, and so on.
}

\change{
Across these different approaches, however, equitable participation cannot be presumed—it must be actively cultivated through attention to power dynamics~\cite{harrington2019deconstructing, delgadoParticipatoryTurnAI2023, ahmed2025ethics}. 
Throughout this project, creating student-centered spaces was not simply a methodological preference but a prerequisite for surfacing concerns that other approaches miss~\cite{puHowCanWe2025}. 
This means attending to who facilitates (student-led or neutral bodies like university libraries may elicit more candid responses than faculty-led effort), how data is collected (anonymous approaches reduce fear of traceability), and whether the process is ongoing. 
On this last point: policies are not one and done. 
The AI landscape changes quickly as new tools and capabilities emerge, and policies wedded to particular platforms or interaction patterns (e.g., chat-based dialogue systems) may become deprecated rapidly. 
It is therefore essential to treat policies as living documents, creating sustained channels of communication with students rather than one-off consultations.
}

\section{Limitations}
This study draws on a single graduate design cohort at a minority-serving public university, limiting generalizability. 
Facilitators seeded policy topics as sensitizing entry points to AI policy design~\cite{sanders2008co}, but this may have induced confirmation bias.
In addition, this study centers student perspectives---by design---and does not incorporate educator or administrator viewpoints.
The resulting policies therefore are not always feasible or reasonable to implement, especially when incongruent with learning objectives. 
Finally, our workshops did not address consequences for policy violations, but rather explored how student participation may lessen the likelihood for violations. 




\change{
\section{Conclusion}
In this paper, we sought student perspectives on AI policies in a design classroom, towards participatory governance of AI in higher education. 
To do so, we positioned students as lead users, or early adopters of generative AI technologies, and looked to their lived experience with these tools as a source of legitimate information to inform policies. 
Through a three-part workshop series, students authored ten policy recommendations, embedded in a zine that has since circulated across campus, sparking discourse around student participation beyond the bounds of the workshops. 
These recommendations surfaced concerns often absent from top-down AI governance—such as the desire for assignment-specific guidance, frustration with perceived double standards in faculty AI use, and support for English learners—though they also raised implementation challenges, particularly when student preferences diverged from learning objectives. 
Finally, we argue that while the policies themselves are specific to a design classroom, the methods for fostering student participation are transferable to other domains. 
We offer strategies for creating channels for student input that instructors across disciplines can adapt to their own contexts. 
As generative AI continues to reshape higher education, we hope this work offers one model for calling students \emph{in} rather than calling students \emph{out}.
}


\begin{acks}
This work was supported by UMBC's Hrabowski Pedagogy Innovation Award. 
First, we thank our participants for their deep and meaningful contributions to this work.
We also that Julie Sayo and Aasmita Bhattacharya for zine guidance and designerly excellence, and Lara Martin for feedback and support. 
Finally, we thanks our reviewers for their encouragement and constructive feedback.
Generative AI tools like ChatGPT and Claude were used to assist with related work searches and summarization of text.
\end{acks}

\bibliographystyle{ACM-Reference-Format}
\bibliography{main}

\appendix
\section{Student-Driven AI Policy Recommendations Zine Pages}
\captionsetup[figure]{justification=centering}
\label{app:appendix}

\renewcommand{\thefigure}{A\arabic{figure}}

\setcounter{figure}{0}

\begin{figure*}[p]
     \centering
     \includegraphics[width=0.95\textwidth]{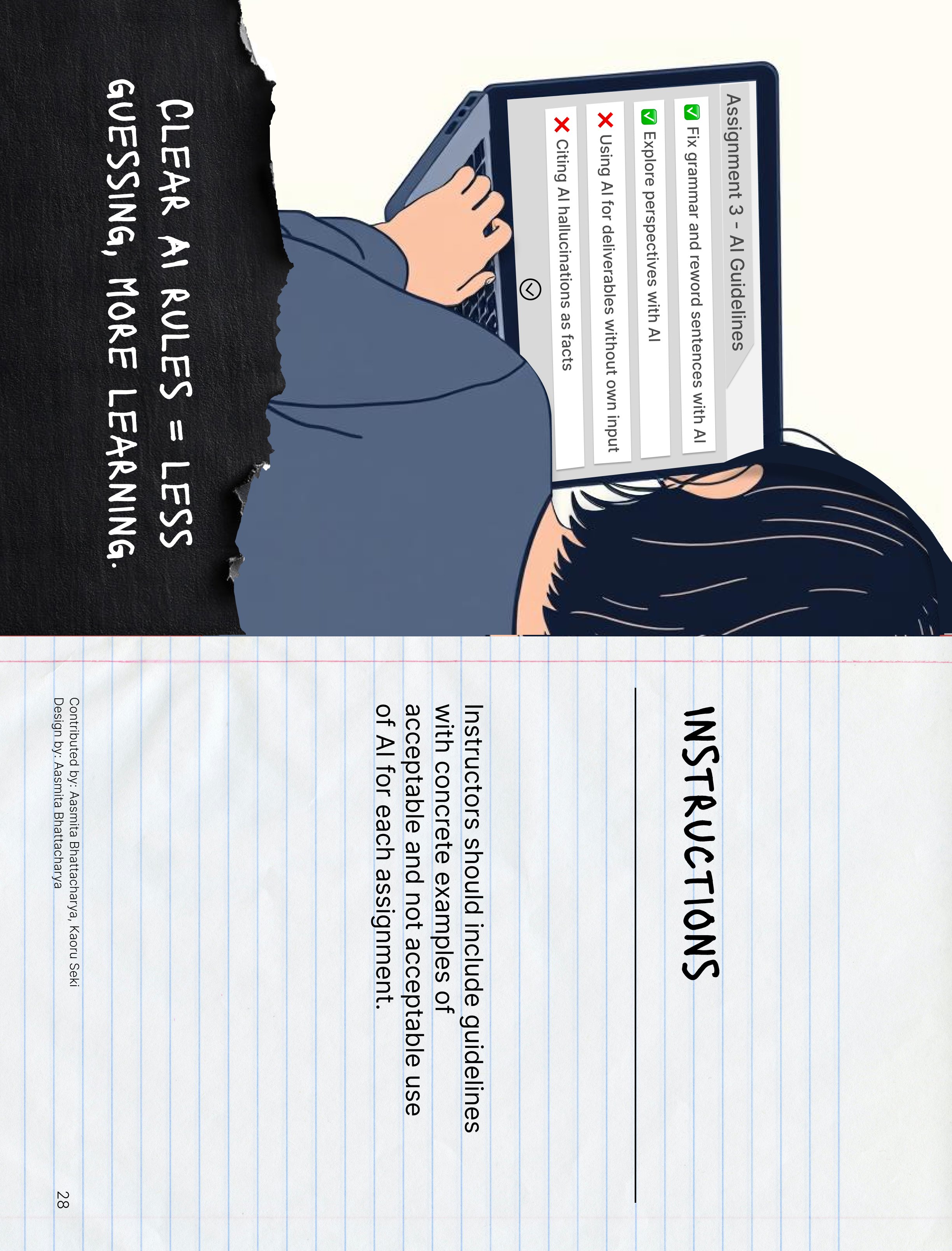}
     \caption{Instructions, Page 1 of 10}
     \Description{On the left, an illustration shows a student looking at a laptop screen displaying “Assignment 3 – AI Guidelines.” The guidelines list acceptable uses with green check marks (such as “Fix grammar and reword sentences with AI” and “Explore perspectives with AI”) and unacceptable uses with red Xs (such as “Using AI for deliverables without our own input” and “Citing AI hallucinations as facts”). Beside the illustration, white text on a black textured background reads: “Clear AI rules = less guessing, more learning.” On the right, a lined notebook paper panel is titled “Instructions” and includes the statement: “Instructors should include guidelines with concrete examples of acceptable and not acceptable use of AI for each assignment.”}
     \label{fig:zine_instruc}
\end{figure*}

\begin{figure*}[p]
     \centering
     \includegraphics[width=0.95\textwidth]{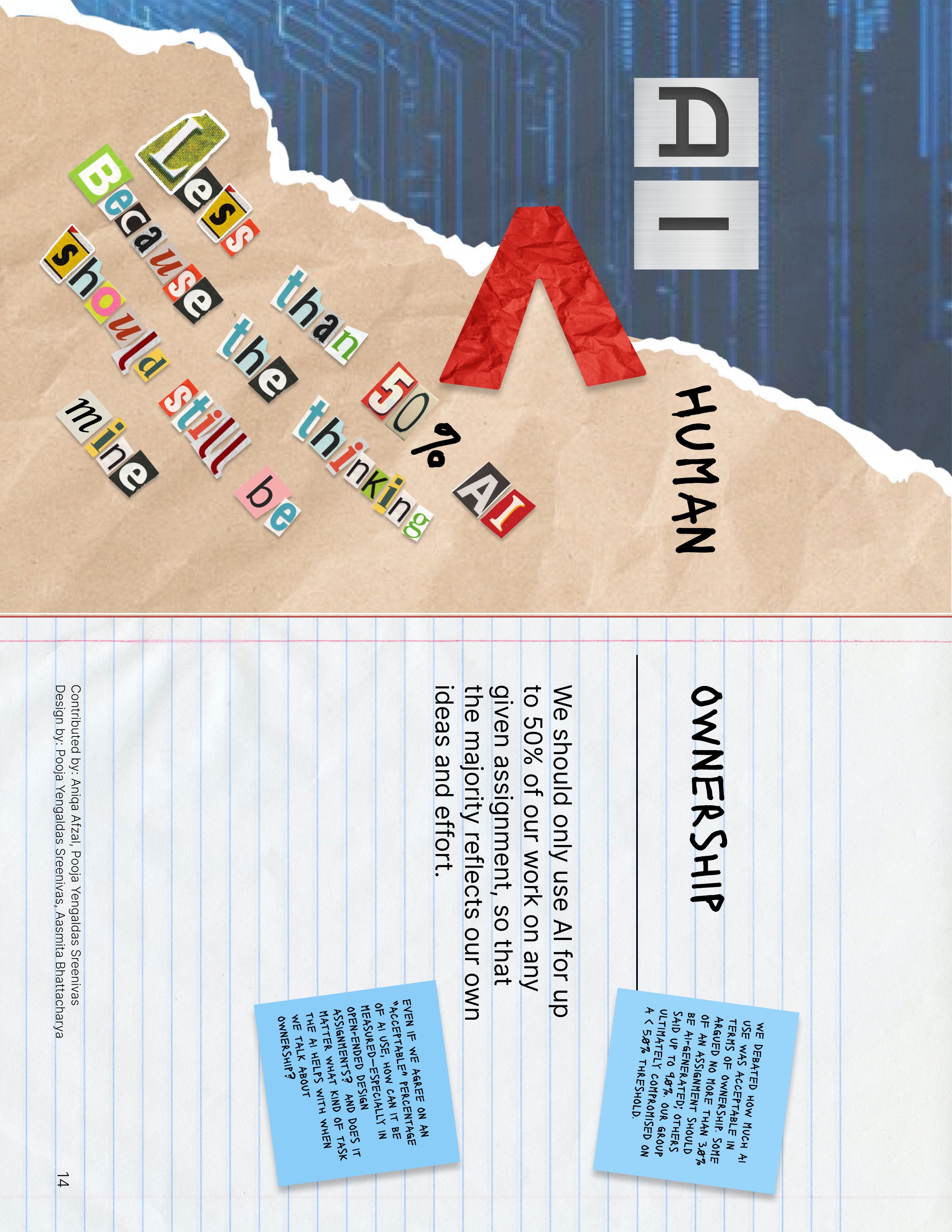}
     \caption{Ownership, Page 2 of 10}
     \Description{The zine spread has two panels. The left panel shows a magazine cutout collage style image that splits in half diagonally in the style of ripped paper. The page reads “AI < Human; Less than 50\% because the thinking should still be mine.” The text is displayed in a found letter style cutout collage. The left half has a black and blue sci-fi style background, and the right side has a beige paper bag style background. The right panel features lined notebook paper with the policy title “OWNERSHIP” with a break line underneath. The statement below reads “We should only use AI for up to 50\% of our work on any given assignment, so that the majority reflects our own ideas and effort.” To the right side of the statement, there are two blue sticky notes. The top one reads “we debated how much AI use was acceptable in terms of ownership. some argued no more than 30\% of an assignment should be AI-generated; others said up to 90\%. our group ultimately compromised on a < 50\% threshold.” The bottom note reads “even if we agree on an “acceptable” percentage of AI use, how can it be measured—especially in open-ended design assignments?  and does it matter what kind of task the AI helps with when we talk about ownership?” }
     \label{fig:zine_owner}
\end{figure*}

\begin{figure*}[p]
     \centering
     \includegraphics[width=0.95\textwidth]{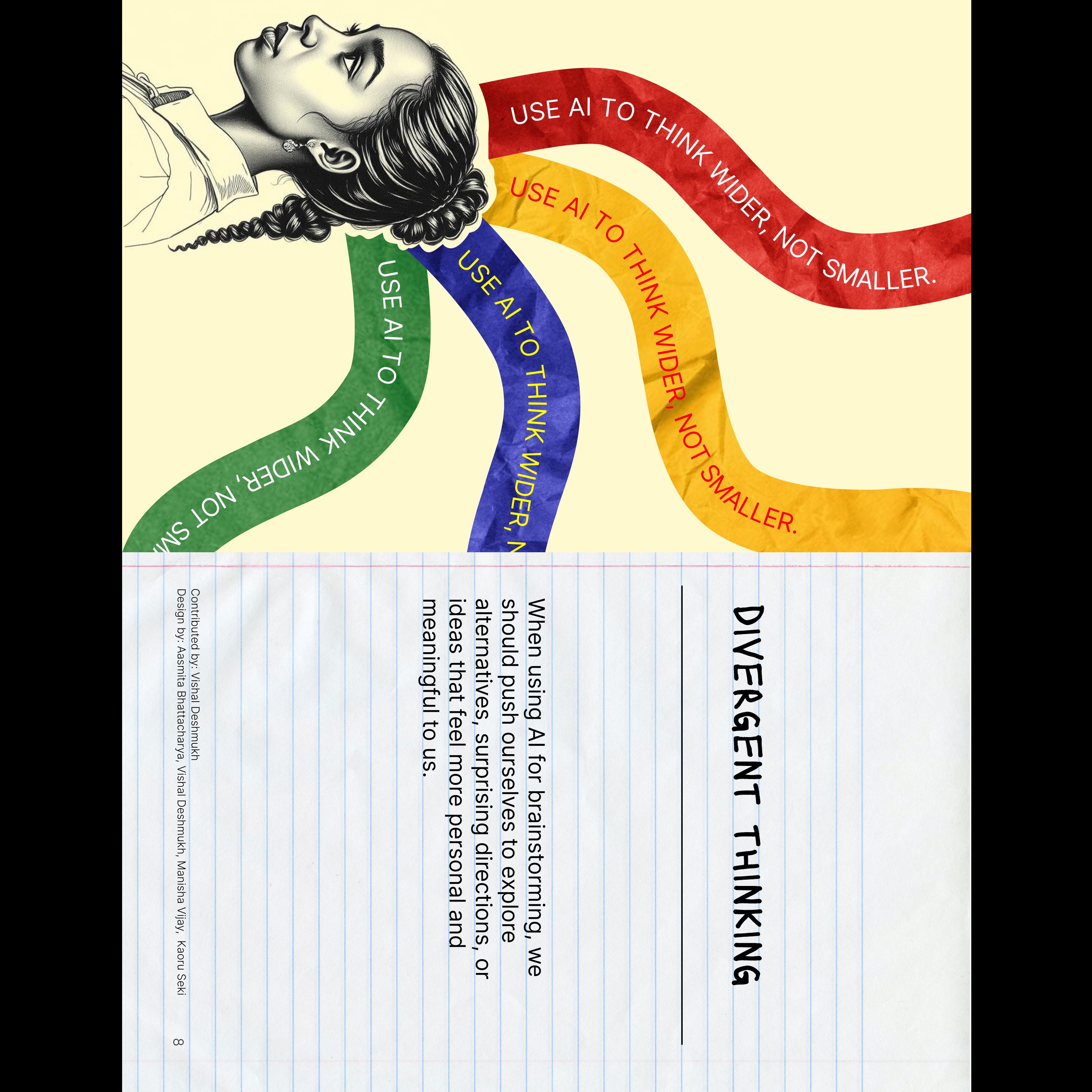}
     \caption{Divergent Thinking, Page 3 of 10}
     \Description{The zine spread has two panels. The left panel shows a stylized illustration of a student in profile, with colorful ribbons of text reading “Use AI to think wider, not smaller,” alongside notebook paper labeled “Divergent Thinking,” which explains that students should use AI for brainstorming in ways that lead to more personal, meaningful ideas. The right panel features lined notebook paper with the text “Divergent Thinking” and an explanation: “If instructors expect us to be transparent about our use of generative AI, we expect the same transparency from them when it comes to how they use AI in teaching, grading, or creating assignments.”}
     \label{fig:zine_diverg}
\end{figure*}

\begin{figure*}[p]
     \centering
     \includegraphics[width=0.95\textwidth]{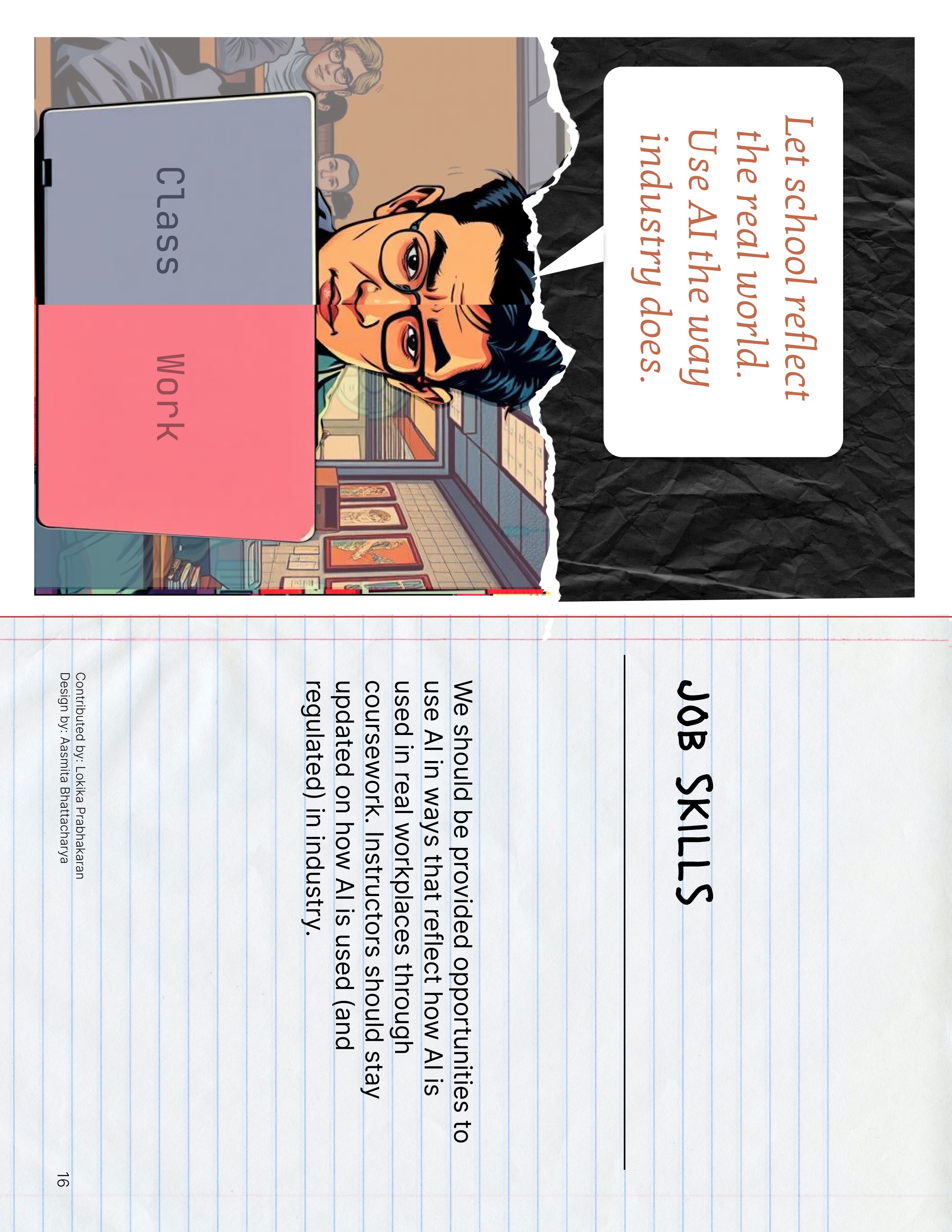}
     \caption{Job Skills, Page 4 of 10}
     \Description{The zine spread has two panels. The left panel is divided over a horizontal paper rip style line. The bottom half features a man with black hair and glasses looking at a laptop, with the back of the laptop split in half to say “class” on the left and “work” on the right. A speech bubble comes above the man’s head reading “Let school reflect the real world. Use AI the way industry does.” The right panel features lined notebook paper with the text “Job Skills” and an explanation: “We should be provided opportunities to use AI in ways that reflect how AI is used in real workplaces through coursework. Instructors should stay updated on how AI is used (and regulated) in industry.” }
     \label{fig:zine_job}
\end{figure*}

\begin{figure*}[p]
     \centering
     \includegraphics[width=0.95\textwidth]{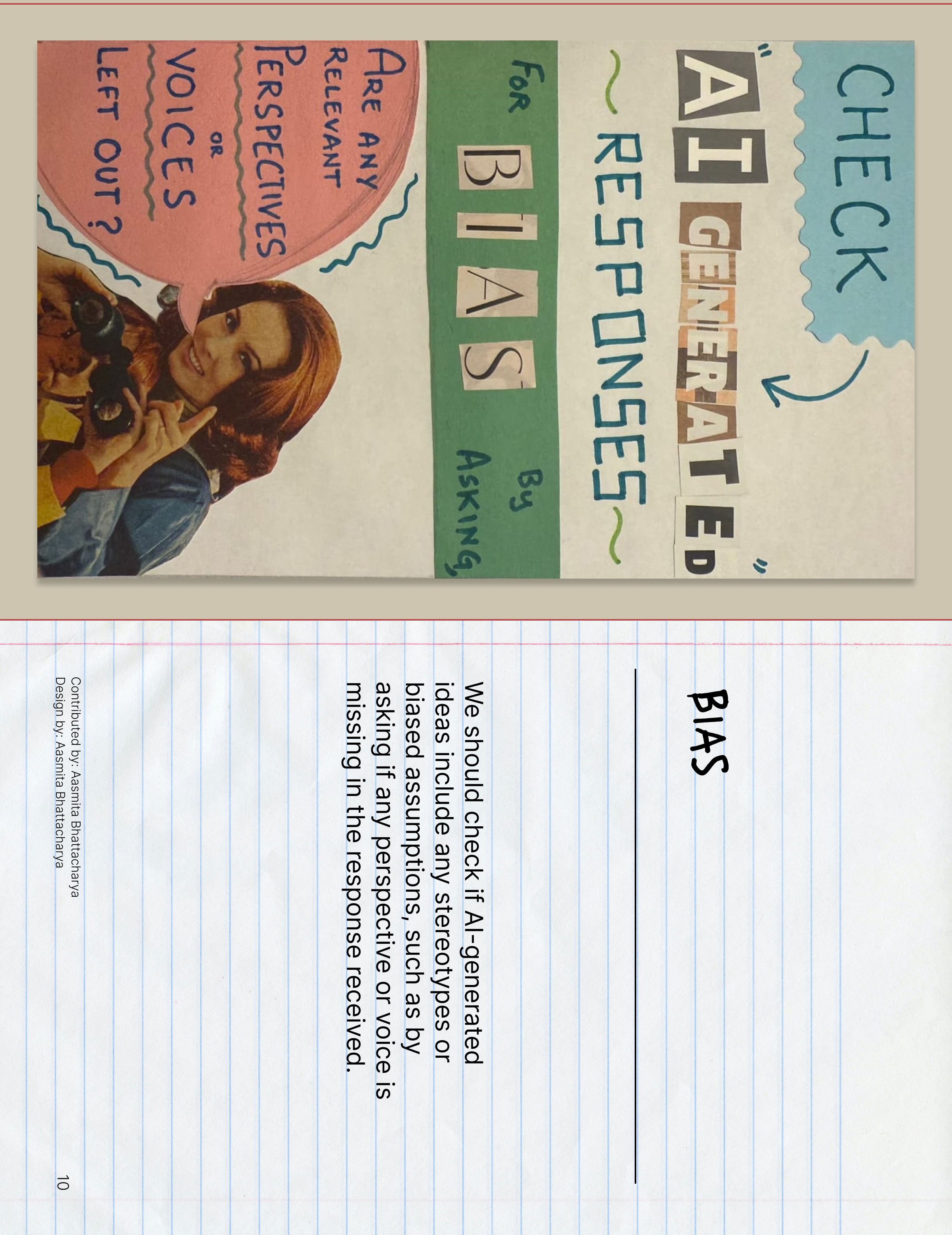}
     \caption{Bias, Page 5 of 10}
     \Description{The zine spread has two panels. The left panel shows a collage style image that reads “Check AI generated responses for BIAS by asking” at the bottom is an image of a woman with shoulder length red hair and a boy standing in front of her with a pair of binoculars. A pink speech bubble comes out of the woman’s mouth finishing the sentence with “Are any relevant perspectives or voices left out?” The right panel features lined notebook paper with the policy title “BIAS” with a heading line, and the statement below reading “We should check if We should check if AI-generated ideas include any stereotypes or biased assumptions, such as by asking if any perspective or voice is missing in the response received.” }
     \label{fig:zine_bias}
\end{figure*}

\begin{figure*}[p]
     \centering
     \includegraphics[width=0.95\textwidth]{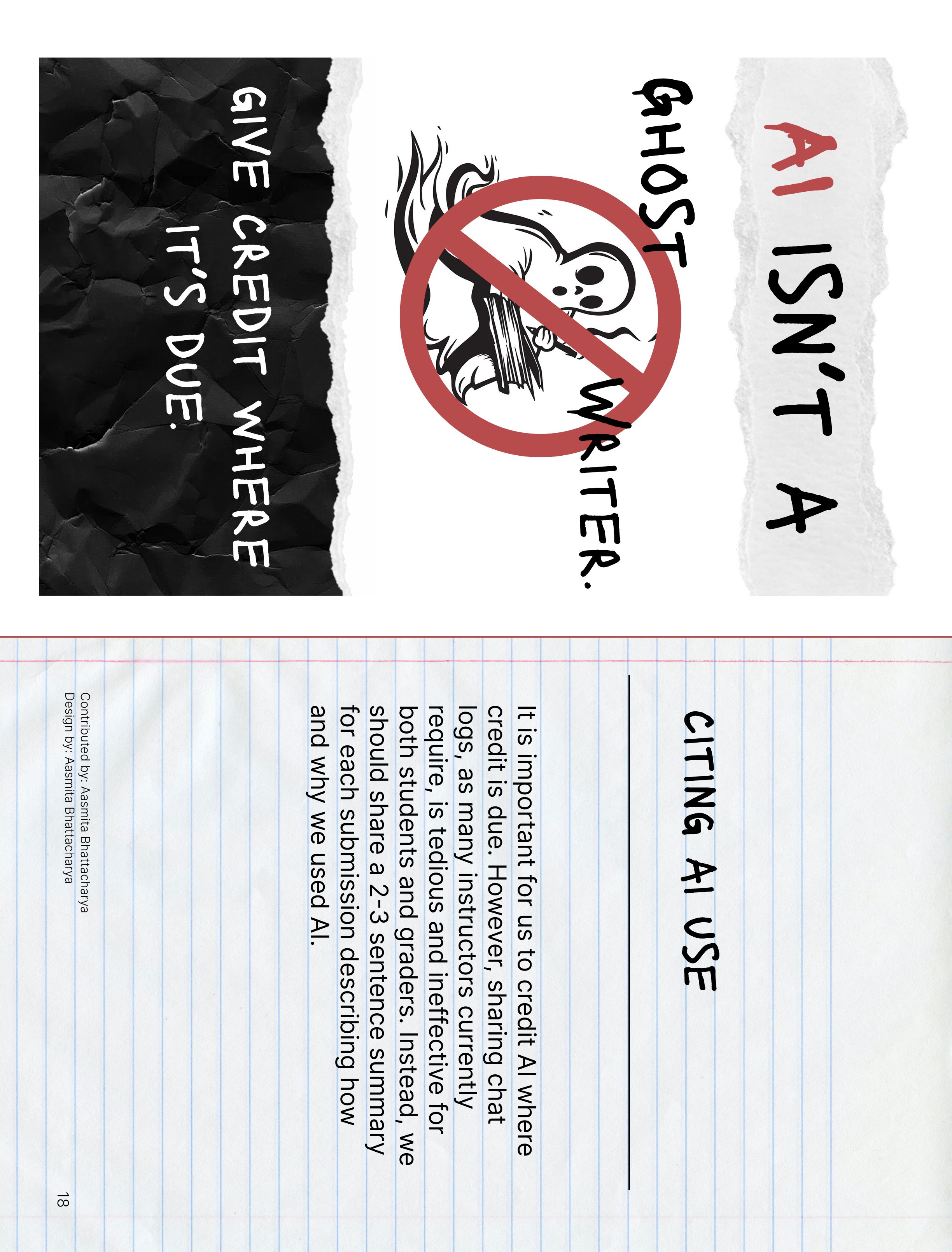}
     \caption{Citing AI Use, Page 6 of 10}
     \Description{The zine spread has two panels. On the left panel, the bottom shows a black textured background with bold white text reading “Give credit where it’s due.” Above it, a cartoon-like ghost holding papers is crossed out with a red prohibition circle, alongside the text “AI isn’t a ghostwriter.” The right panel features lined notebook paper with the text “Citing AI Use” and an explanation: “It is important for us to credit AI where credit is due. However, sharing chat logs, as many instructors currently require, is tedious and ineffective for both students and graders. Instead, we should share a 2-3 sentence summary for each submission describing how and why we used AI.” }
     \label{fig:zine_citing}
\end{figure*}

\begin{figure*}[p]
     \centering
     \includegraphics[width=0.95\textwidth]{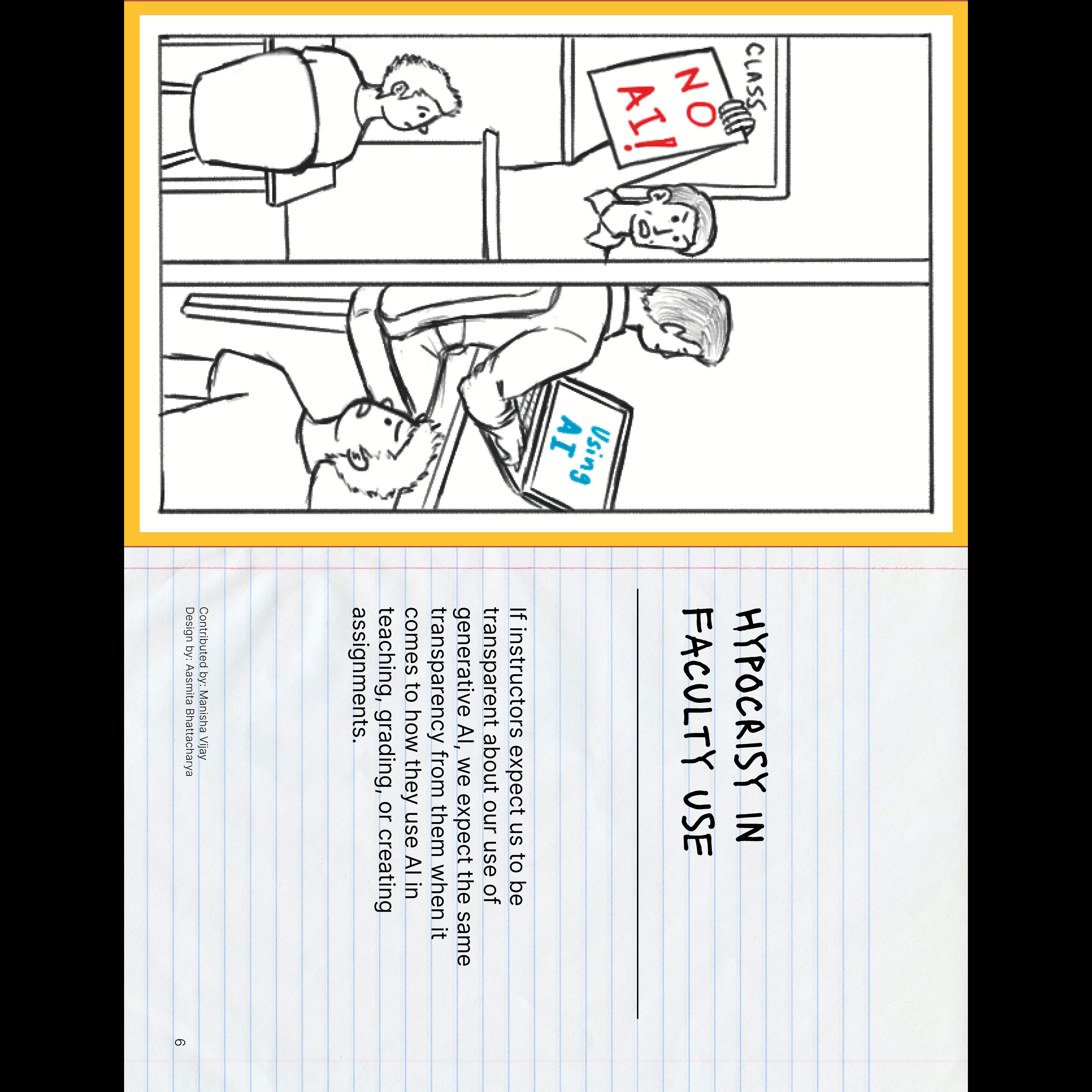}
     \caption{Hypocrisy in Faculty Use, Page 7 of 10}
     \Description{The zine spread has two panels. The left panel is a black-and-white cartoon of a classroom: one student holds a sign reading “No AI!,” while another uses a laptop labeled “Using AI.” The right panel features lined notebook paper with the text “Hypocrisy in Faculty Use” and an explanation: “When using AI for brainstorming, we should push ourselves to explore alternatives, surprising directions, or ideas that feel more personal and meaningful to us.” }
     \label{fig:zine_hypoc}
\end{figure*}

\begin{figure*}[p]
     \centering
     \includegraphics[width=0.95\textwidth]{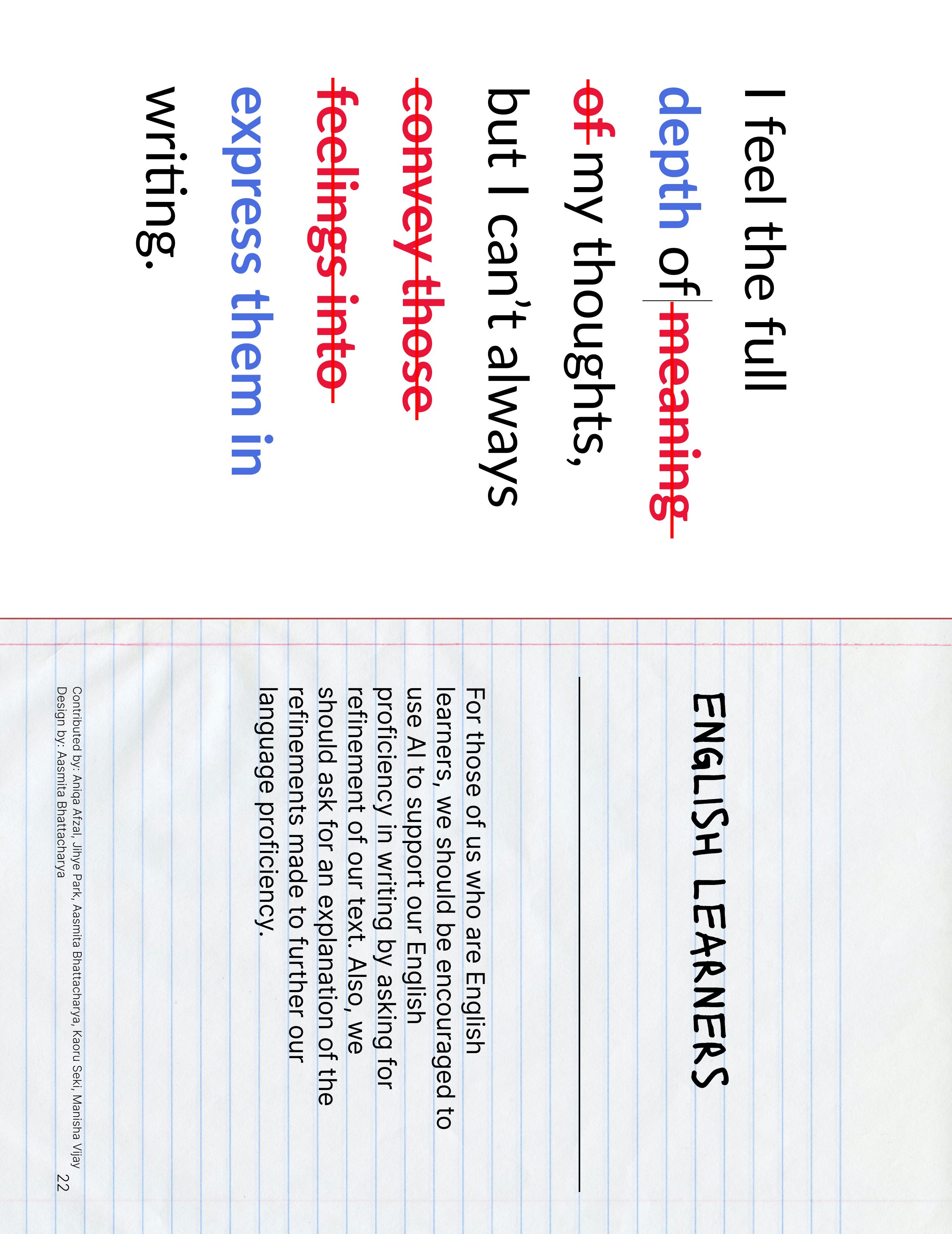}
     \caption{English Learners, Page 8 of 10}
     \Description{The zine spread has two panels. On the left, large stylized text reads: “I feel the full depth of my thoughts, but I can’t always express them in writing.” Some words are crossed out in red (such as “meaning” and “convey those feelings into”) and replaced with alternatives in blue (such as “depth” and “express them in”), visually illustrating the difficulty of articulation in writing. Together, the two panels convey how AI can help English learners refine their writing while deepening language learning. On the right, a panel of lined notebook paper is labeled “English Learners” and includes the statement: “For those of us who are English learners, we should be encouraged to use AI to support our English proficiency in writing by asking for refinement of our text. We should also ask for an explanation of the refinements made to further our language proficiency.” }
     \label{fig:zine_english}
\end{figure*}

\begin{figure*}[p]
     \centering
     \includegraphics[width=0.95\textwidth]{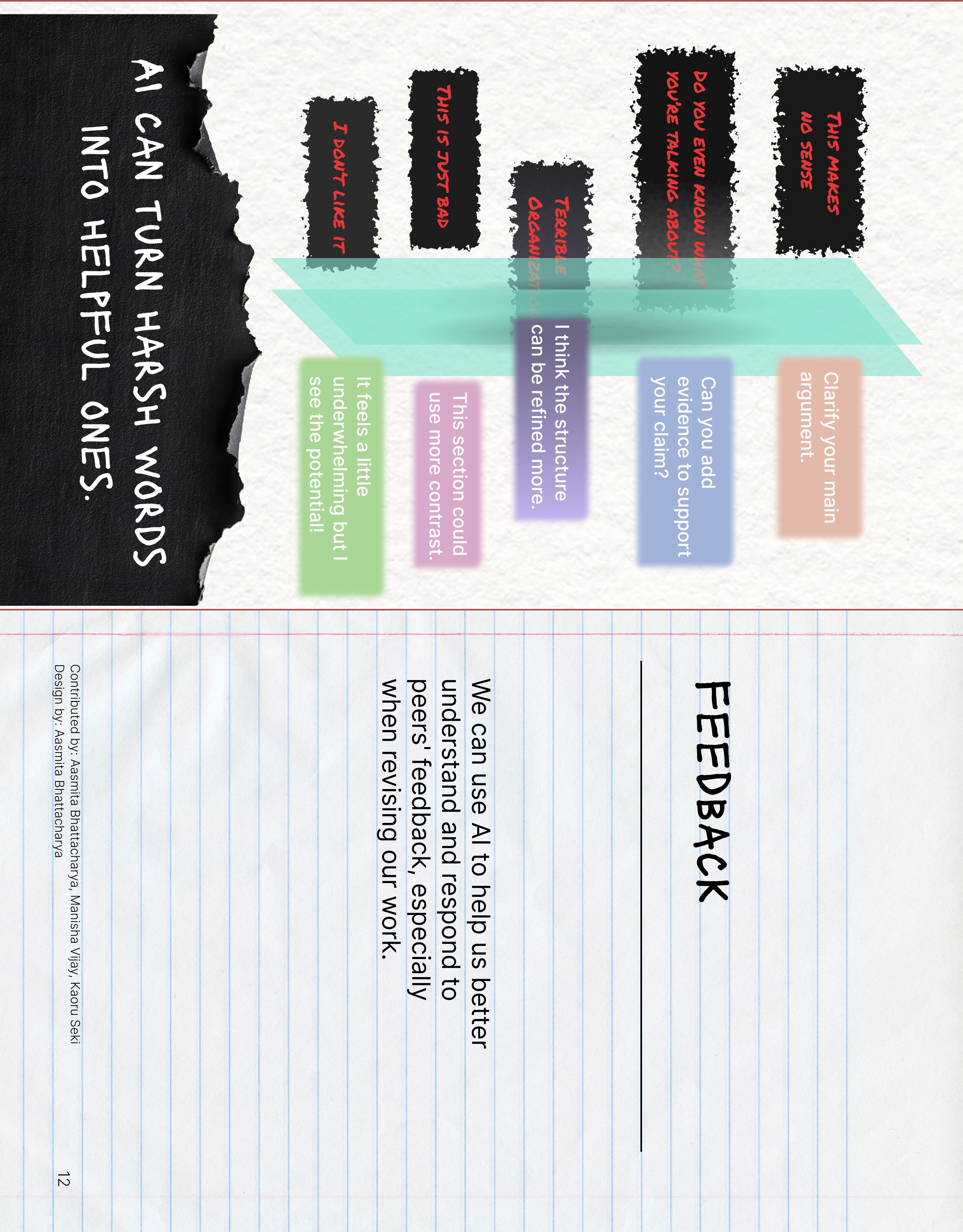}
     \caption{Feedback, Page 9 of 10}
     \Description{The zine spread has two panels. On the left, white text on a black textured background reads: “AI can turn harsh words into helpful ones.” Above this phrase, several harsh feedback comments appear in red and black, such as “I don’t like it,” “This is just bad,” “Do you even know what you’re talking about?,” and “This makes no sense.” Passing through the blue filter, these comments are transformed into constructive phrases displayed in colorful boxes, such as “It feels a little underwhelming but I see the potential,” “This section could use more contrast,” “I think the structure can be refined more,” and “Can you add evidence to support your claim?” The right panel features lined notebook paper with the text “Feedback” and an explanation: “We can use AI to help us better understand and respond to peers' feedback, especially when revising our work.” }
     \label{fig:zine_feedback}
\end{figure*}

\begin{figure*}[p]
     \centering
     \includegraphics[width=0.95\textwidth]{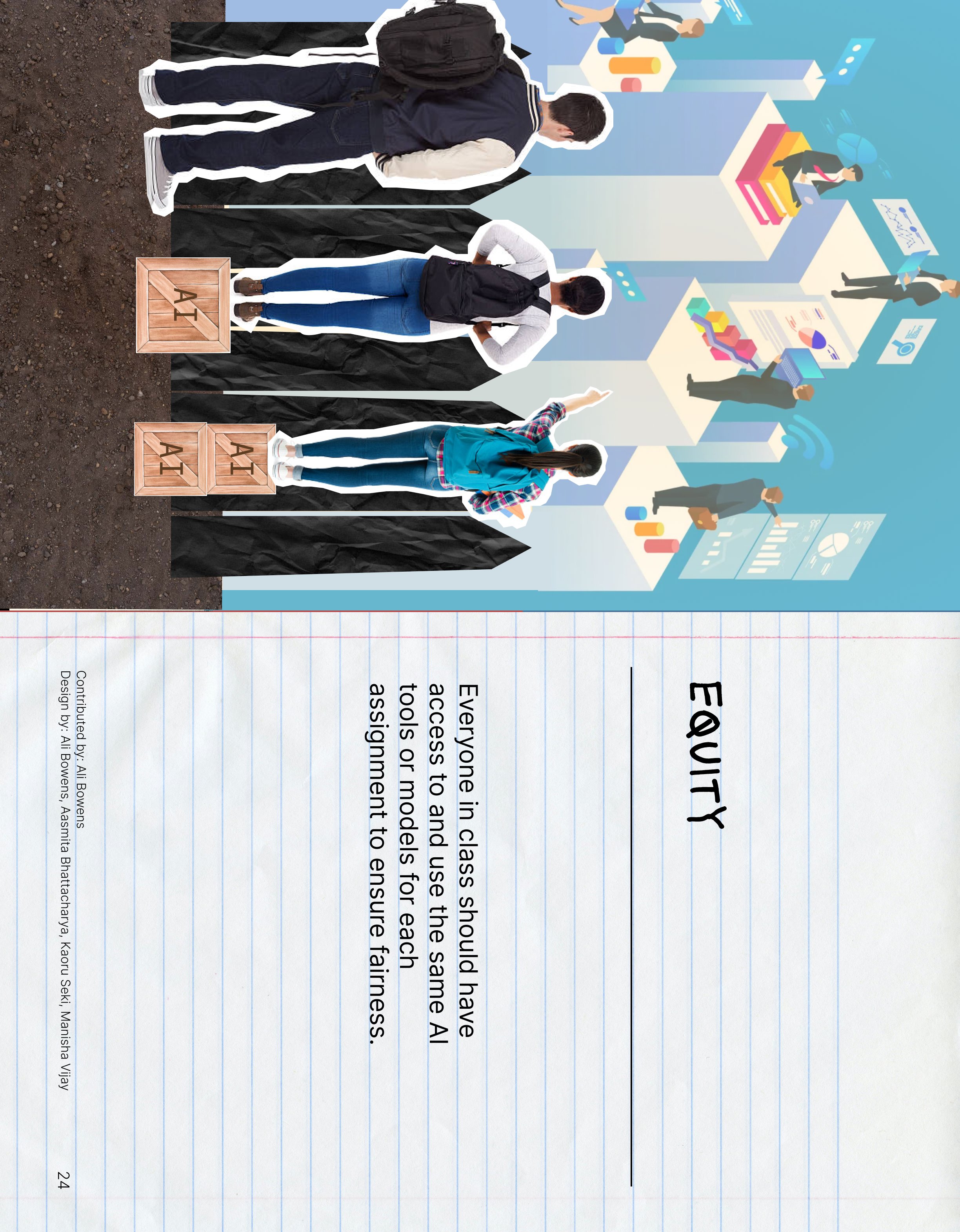}
     \caption{Equity, Page 10 of 10}
     \Description{The zine spread has two panels. On the left, a collage illustration depicts three students facing forward toward a futuristic workplace scene with charts, books, and office workers. The students stand on uneven black platforms: one has no support, another stands on a single wooden crate labeled “AI,” and the third on two crates stacked together. This imagery highlights inequity in access. On the right, a lined notebook paper panel is titled “Equity” and includes the statement: “Everyone in class should have access to and use the same AI tools or models for each assignment to ensure fairness.”}
     \label{fig:zine_equity}
\end{figure*}

\end{document}